\input harvmac 
\input epsf.tex
\def\IN{\relax{\rm I\kern-.18em N}} 
\def\IR{
\relax{\rm I\kern-.18em R}} \font\cmss=cmss10 
\font\cmsss=cmss10 at 7pt \def\IZ{\relax\ifmmode\mathchoice 
{\hbox{\cmss Z\kern-.4em Z}}{\hbox{\cmss Z\kern-.4em Z}} 
{\lower.9pt\hbox{\cmsss Z\kern-.4em Z}} {\lower1.2pt\hbox{
\cmsss Z\kern-.4em Z}}
\else{\cmss Z\kern-.4em Z}\fi} 

\def\deb{{\par
\vbox{
$ \ \ \ \ \ \ \ \ \ \ \ \ \ \ \ \ \ \ \ \ \ \ \ \ \ \ \ \ \ \ \ \
\ \ \ \ \ \ \ \ \ \ \ \ \ \ \ \ \ \ \ \ \ \ \ \ \ \ \ \ \ \ \ \
\ \ \ \ \ \ \ \ \ \ \ \ \ \ \ \ \ \ \ \ \ \ \ \ \ \ \ \ \ \ \ \ $
\par
\hrule}\par}}
\def\fin{{\par
\vbox{
\hrule \par
$ \ \ \ \ \ \ \ \ \ \ \ \ \ \ \ \ \ \ \ \ \ \ \ \ \ \ \ \ \ \ \ \
\ \ \ \ \ \ \ \ \ \ \ \ \ \ \ \ \ \ \ \ \ \ \ \ \ \ \ \ \ \ \ \
\ \ \ \ \ \ \ \ \ \ \ \ \ \ \ \ \ \ \ \ \ \ \ \ \ \ \ \ \ \ \ \ $
}\par }}
\overfullrule=0mm
\def\file#1{#1}
\newcount\figno \figno=0
\newcount\figtotno      
\figtotno=0
\newdimen\captionindent 
\captionindent=1cm 
\def\figbox#1#2{\epsfxsize=#1\vcenter{
\epsfbox{\file{#2}}}} 
\newcount\figno
\figno=0
\def\fig#1#2#3{ \par\begingroup\parindent=0pt
\leftskip=1cm\rightskip=1cm\parindent =0pt 
\baselineskip=11pt
\global\advance\figno by 1
\midinsert
\epsfxsize=#3
\centerline{\epsfbox{#2}}
\vskip 12pt
{\bf Fig. \the\figno:} #1\par
\endinsert\endgroup\par
}
\def\figlabel#1{\xdef#1{\the\figno}} 
\def\encadremath#1{\vbox{\hrule\hbox{\vrule\kern8pt 
\vbox{\kern8pt \hbox{$\displaystyle #1$}\kern8pt} 
\kern8pt\vrule}\hrule}} \def\enca#1{\vbox{\hrule\hbox{
\vrule\kern8pt\vbox{\kern8pt \hbox{$\displaystyle #1$}
\kern8pt} \kern8pt\vrule}\hrule}}

\def\IR{\relax{\rm I\kern-.18em R}}
\font\cmss=cmss10 \font\cmsss=cmss10 at 7pt 
\def\IZ{\relax\ifmmode\mathchoice
{\hbox{\cmss Z\kern-.4em Z}}{\hbox{\cmss Z\kern-.4em Z}} 
{\lower.9pt\hbox{\cmsss Z\kern-.4em Z}}
{\lower1.2pt\hbox{\cmsss Z\kern-.4em Z}} 
\else{\cmss Z\kern-.4em Z}\fi} \def\buildrel#1\under#2{ 
\mathrel{\mathop{\kern0pt #2}\limits_{#1}}}
\Title{UNC-CH/1996}
{{\vbox {
\bigskip
\centerline{Meander Determinants}
}}}
\bigskip
\centerline{P. Di Francesco\footnote*{e-mail: philippe@math.unc.edu},}
\bigskip
\centerline{Department of Mathematics,} 
\centerline{\it University of North Carolina at Chapel Hill,} 
\centerline{\it  CHAPEL HILL, N.C. 27599-3250, U.S.A.} 
\centerline{and}
\centerline{ \it Service de Physique Th\'eorique,} 
\centerline{C.E.A. Saclay,}
\centerline{ \it F-91191 Gif sur Yvette Cedex, France} 
\vskip .5in
\noindent
We prove a determinantal formula for quantities related to the 
problem of enumeration of (semi-) meanders, namely
the topologically inequivalent
planar configurations of non-self-intersecting loops
crossing a given
(half-) line through a given number of points.
This is done by the explicit Gram-Schmidt orthogonalization
of certain bases of subspaces of the Temperley-Lieb algebra. 
\Date{11/96}
\nref\HMRT{K. Hoffman, K. Mehlhorn, P. Rosenstiehl and
R. Tarjan, {\it Sorting Jordan sequences in linear time 
using level-linked 
search trees}, Information and Control {\bf 68} (1986) 170-184.} 
\nref\TOU{J. Touchard, 
{\it Contributions \`a l'\'etude du probl\`eme des timbres poste}, 
Canad. J. Math. {\bf 2} (1950) 385-398.}
\nref\LUN{W. Lunnon, {\it A map--folding problem}, 
Math. of Computation {\bf 22} 
(1968) 193-199.}
\nref\PHI{A. Phillips, 
{\it La topologia dei labirinti}, in M. Emmer, ed. 
{\it L' occhio di Horus: itinerario nell'immaginario matematico}, 
Istituto della 
Enciclopedia Italia, Roma (1989) 57-67.}
\nref\ARNO{V. Arnold, {\it The branched covering of $CP_2 \to S_4$, 
hyperbolicity and projective topology}, 
Siberian Math. Jour. {\bf 29} (1988) 717-726.} 
\nref\KOSMO{K.H. Ko, L. Smolinsky, {\it A combinatorial matrix 
in $3$-manifold theory}, 
Pacific. J. Math {\bf 149} (1991) 319-336.} 
\nref\LZ{S. Lando and A. Zvonkin, 
{\it Plane and Projective Meanders}, Theor. Comp. 
Science {\bf 117} (1993) 227-241, 
and {\it Meanders}, Selecta Math. Sov. {\bf 11} (1992) 117-144.} 
\nref\DGG{P. Di Francesco, O. Golinelli 
and E. Guitter, {\it Meander,
folding and arch statistics}, 
to appear in Mathematical and Computer 
Modelling (1996), hep-th/950630.}
\nref\MAK{Y. Makeenko, {\it Strings, Matrix Models and Meanders}, 
proceedings of 
the 29th Inter. Ahrenshoop Symp., Germany (1995).}
\nref\DGGB{P. Di Francesco, O. Golinelli and E. Guitter, {\it 
Meanders and the Temperley-Lieb algebra}, to appear in Comm. Math. Phys.
(1996), hep-th/9602025.}
\nref\TLA{H. Temperley and E. Lieb, 
{\it Relations between the percolation and coloring problem and 
other graph-theoretical problems associated with regular planar 
lattices: some exact results for the percolation problem}, 
Proc. Roy. Soc. {\bf A322} 
(1971) 251-280.}
\nref\MARTIN{P. Martin, 
{\it Potts models and related problems in statistical mechanics}, 
World Scientific (1991).}
\nref\DIF{P. Di Francesco, {\it Integrable lattice models, graphs
and modular invariant conformal field theories}, Int. Jour. Mod.
Phys. {\bf A7} (1992) 407-500.}
\nref\RESHE{N. Reshetikhin, {\it Quantized universal 
enveloping algebras, the Yang-Baxter equation and invariants of links 
1 and 2}, LOMI preprints E-4-87 and E-17-87 (1988).}
\newsec{Introduction}
The meander problem  
consists in counting the number 
$M_n$ of meanders of order $n$, i.e. of 
inequivalent configurations
of a closed non-self-intersecting
loop crossing an infinite line through $2n$ points.  
The infinite line may be
viewed as a river flowing from east to west, and the 
loop as a closed circuit crossing this river through $2n$ bridges,
hence the name ``meander", although here the river and the road play
symmetric roles.  
Two configurations are considered as equivalent if they 
are smooth deformations of one another. 
The meander problem probably first arose in the work of 
Poincar\'e about differential geometry. Since then, 
it has emerged in 
many different contexts, such as mathematics, physics, 
computer science \HMRT\ \TOU\ \LUN\
and even fine arts \PHI. 
The problem was recently reactualized by Arnold, 
in relation with Hilbert's 16th 
problem \ARNO. 
Meanders also emerged in the classification 
of 3-manifolds \KOSMO.  More recently, random 
matrix model techniques, borrowed from quantum 
field theory, 
were applied to this problem \LZ\ \DGG\ \MAK. \par
In the present paper, we rather adopt the purely algebraic approach
advocated in \DGGB, based on a pictorial representation of 
the elements of the Temperley-Lieb algebra 
\TLA\ 
(see also P. Martin's book
\MARTIN\ for an elementary introduction) 
using strings
(each element is a sort of domino with string-ends on its boundary,
and elements are multiplied by connecting the string-ends of the
corresponding dominos), by
means of which the meanders are constructed.
A particular
basis (set of basic dominos) of the Temperley-Lieb algebra
will provide us with the building blocks for the construction of
meanders, or some of their generalizations, the semi-meanders,
introduced in \DGG.
Roughly speaking, a (multi-component, i.e.
made of possibly several non-intersecting roads) meander is obtained as the 
concatenation of two dominos, and the identification of their
free string-ends: this is exactly the manipulation involved
when evaluating the standard bilinear form of the Temperley-Lieb
algebra on these two dominos. 
Moreover, the value of the bilinear form
is simply
$ q^L$, $q$ a given complex number, and $L$ the total number
of loops formed by the connection of the strings, namely the number
of loops in the corresponding meander.  
The Theorem 1 below is a formula expressing the determinant
of the Gram matrix of this basis of the Temperley-Lieb algebra,
and was first derived in \DGGB\ (an algorithm for its computation was
also given in \KOSMO).
\par
By choosing 
a particular subset of dominos (hence a particular
basis of some subspace of the Temperley-Lieb algebra), 
we may obtain meanders with more
specific details: the semi-meanders with fixed winding numbers 
are some of these. 
To obtain the latter, consider the following semi-meander problem:
enumerate the inequivalent planar configurations of 
a loop crossing a {\it half-line} through a given number of points $n$.
The loop of such a semi-meander may freely wind around the origin of
the half-line (interpreted as the source of the river), and
we can define a winding number associated to this. The (multi-component,
i.e. with possibly several non-intersecting roads) semi-meanders
with fixed winding number $w$ may be obtained as the concatenation
of particular elements (dominos) of the Temperley-Lieb algebra,
namely those with  exactly $w$ of the $n$ strings going across the
domino. 
The Theorem 2 below is the generalization to semi-meanders of the
abovementioned meander determinant formula. 
The latter was only conjectured in \DGGB, in a slightly different form.
\par
\medskip
The paper is organized as follows. 
In Sect.2, we recall a few definitions and facts on 
(semi-) meanders, in particular their various formulations as
(i) superpositions of two (open)  
arch configurations (ii) superposition of two (open) walk diagrams,
and we state the main results of the paper (Theorems 1 and 2), in
the form of determinant formulas.
\par
Sect.3 is devoted to the proof of the formula for the meander
determinant. The proof relies on the interpretation of any meander
as the product of two elements of the Temperley-Lieb
algebra. After displaying the various mappings between 
arch configurations, walk diagrams and reduced elements of the
Temperley-Lieb algebra, we reformulate the meander determinant as 
the Gram determinant of a particular basis of the Temperley-Lieb
algebra,
or rather of one of its ideals. The proof is then carried out, by 
performing the explicit Gram-Schmidt orthogonalization of this basis
(Proposition 1). This appears in fact as the consequence of a stronger
statement regarding the orthogonalization of all the {\it products}
of any two basis elements (Lemma 1). 
The computation of the meander determinant is then a combinatorial
exercise (Proposition 2) in rearranging all the normalization
factors introduced in the orthonormalization process, which we carry out
by performing some mapping of decorated walk diagrams.
\par
In Sect.4, we turn to the semi-meander generalization.
We follow the same strategy as in Sect.3, with a number of
complications, due to the fact we now deal with a {\it subspace} of the
Temperley-Lieb algebra, which is not an ideal, i.e. has no good
multiplication properties between its elements. Nevertheless,
we are still able to perform the explicit Gram-Schmidt orthogonalization
of our initial basis (Proposition 3 and Lemmas 2, 3, 4). 
The determinant formula then follows from a rearrangement 
(Propositions 4, 5 and 6) of the normalization factors introduced
in the orthonormalization process. 
\par
We gather in Sect.5 a few
concluding remarks.
\par
\newsec{Meander determinants: the results}
\subsec{Arch configurations and meanders}
A {\it meander} of order\foot{In this paper, the order will 
always refer to the total number of bridges in the configuration. 
A different convention was adopted in refs. \DGG\ \DGGB, where the 
order of a meander is rather half its number of bridges.}
$2n$ is a planar configuration of a closed non-selfintersecting 
loop (road) crossing a line (river) through $2n$ distinct 
points (bridges), considered up to 
smooth deformations preserving the topology of the 
configuration (i.e., preserving the succession of bridges). 
\par
\fig{Any meander is obtained as the superimposition of a top
($a$) and bottom ($b$) arch configurations of same order 
($2n=10$ here).
An arch configuration is a planar pairing
of the $(2n)$ bridges through $n$ non-intersecting arches
lying above the river (by convention, we represent
the lower configuration $b$ reflected with respect to the
river: this will actually be denoted by $b^t$ in the 
following).}{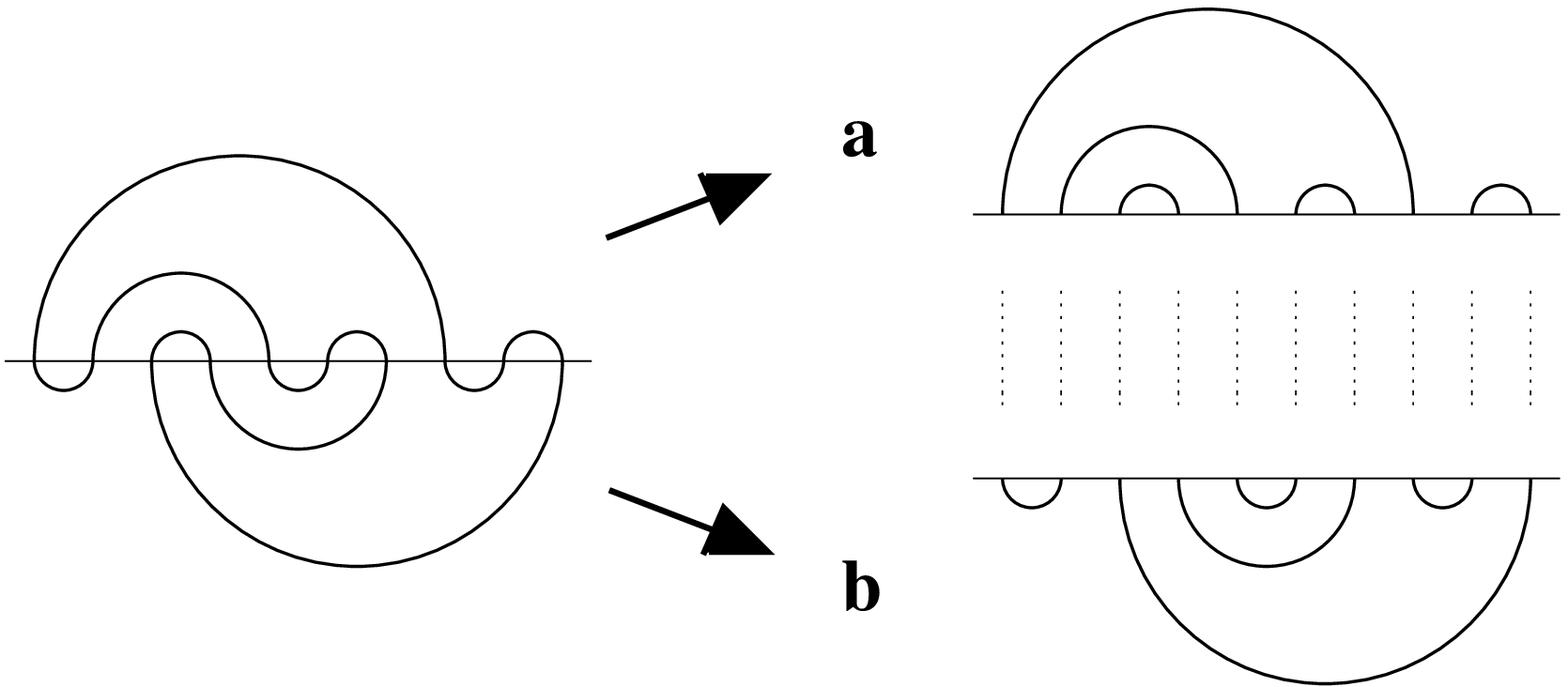}{10.cm}
\figlabel\archm
The river separates the meander into an upper and a lower 
planar configuration of $n$ non-intersecting
pieces of road (arches) joining the $2n$ bridges 
by pairs (see Fig.\archm\ for an example), 
respectively contained in the upper and lower 
half-planes defined by the river. Such a configuration, 
considered up to the abovementioned equivalence, 
is called an {\it arch configuration} 
of order $2n$. Let $A_{2n}$ denote the set of all 
arch configurations 
of order $2n$.
\par
Let us label the bridges from left to right $1,2,...,2n$.
The total number $c_n$ of arch configurations of order $n$ is obtained
by considering the leftmost arch, joining say the bridge 1 to the
bridge $2j$, $1 \leq j \leq n$. This arch separates the arch 
configuration into two pieces: the portion below the 
leftmost arch, and that to the right of this arch. 
These are two arbitrary arch configurations of 
respective orders $2(j-1)$ and $2(n-j)$. Hence we have 
$c_n=\sum_{1\leq j \leq n} c_{j-1} c_{n-j}$. With $c_0=1$, 
we get 
\eqn\cata{ |A_{2n}|~=~ c_n~=~{(2n)! \over (n+1)! n!} }
which is the Catalan number of order $n$.
\par
Superposing two arbitrary arch configurations $a,b \in A_{2n}$ (after
a reflection of $b$ w.r.t. the river)
will
in general lead to a multi-component meander, made of several 
non-intersecting roads. We denote by $\kappa(a|b)$ 
the corresponding number of 
connected components. 
\par
\subsec{Open arch configurations and semi-meanders}
A semi-meander of order $n$ is a planar configuration of a 
non-selfintersecting loop (road) crossing a half-line
(river with a source) through $n$ distinct points (bridges),
up to smooth deformations of the road preserving the topology of
the configuration. 
The main difference with a meander is that the road may now freely
wind around the source of the river, therefore
the number of bridges needs not be an even integer. 
We define the winding number of
a semi-meander to be the number of pairs of bridges linked 
by an arch encircling 
the source of the river (each such arch contributes 1 to the 
total winding number). Note that a semi-meander of order $n$ 
may only have 
a winding number $h=n$ mod 2.
\par
\fig{Any semi-meander may be viewed as the superimposition of an upper
and a lower
open arch configurations. Here the initial semi-meander has
order $n=5$ and winding $h=3$.
The two open arch configurations on the right have $h=3$ open arches. To
recover the initial semi-meander, these open arches must be connected
two by two,
from the right to the left
(the arches number 5,4,3 of the upper configuration are respectively
connected to
the arches number 3,2,1 of the lower
configuration).}{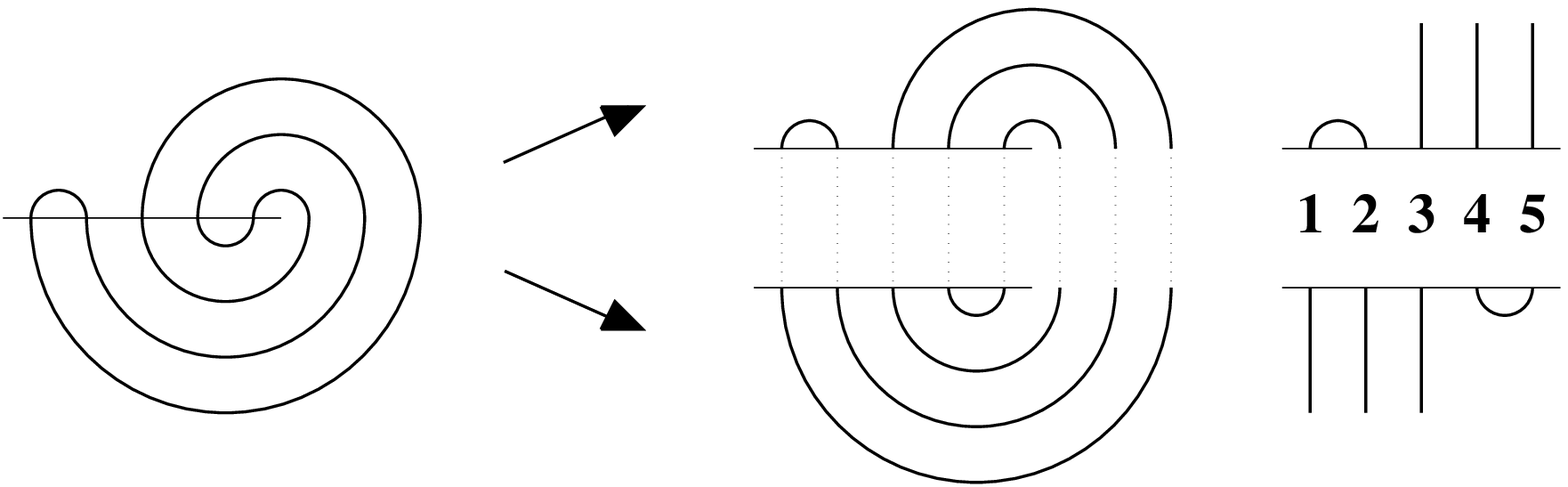}{10.cm}
\figlabel\smopen
In any given semi-meander with winding number $h$, the river still 
separates the configuration into an upper and a lower one (see
Fig.\smopen\ for an example), 
corresponding to the portion lying respectively above and 
below the river, but these 
are linked by $h$ arches encircling the source of the river, 
and connecting $h$ upper parts of bridges to $h$ lower parts 
of bridges. Let us cut these $h$ arches and extend them so 
as to form $h$ vertical half-lines
on the upper configuration and $h$ vertical 
half-lines on the lower 
configuration. The resulting objects 
are called {\it open arch 
configurations} of order $n$ with $h$ open arches\foot{As before,
the order refers to the total number of bridges in the 
configuration.}. Such an open arch configuration is formed
by a line with $n$ distinct points (upper half-bridges) 
either connected by pairs through arches in the upper half-plane (there
are $(n-h)/2$ such arches), or
connected to ``infinity" through a vertical half-line (there are 
$h$ such open arches), with a total of $(n+h)/2$ arches.
We denote by $A_n^{(h)}$ the set of open arch
configurations of order $n$ with $h$ open arches. 
Note again that $h=n$ mod 2.
In particular, $A_{2n}^{(0)}=A_{2n}$.
\par
To compute the cardinal $c_{n,h}$
of $A_n^{(h)}$, we concentrate again on the
leftmost arch of a given configuration. Two cases may occur: 
\item{(i)} This arch is open. The configuration lying on the 
right of this arch is an arbitrary
open arch configuration of order $n-1$ with $h-1$ 
open arches.
\item{(ii)} This arch connects the 
bridges 1 and say $2j$, thus separating the 
configuration into two parts: the one below the leftmost arch 
is an arbitrary arch configuration of order $2(j-1)$, 
whereas the one to the right of bridge 
$2j$ is an arbitrary open arch configuration of order $n-2j$ with $h$ 
open arches.
\par
These are summarized in the following recursion relation 
\eqn\recopar{ c_{n,h}~=~c_{n-1,h-1}+\sum_{j=1}^{[n/2]} 
c_{j-1} c_{n-2j,h} }
where $c_n$ denotes the Catalan number \cata, and $[x]$ is 
the largest integer smaller or equal to $x$.
With the initial condition $c_{2n,0}=c_n$ for
all $n\geq 0$, this determines the numbers $c_{n,h}$ completely, 
and we have
\eqn\oparnum{ |A_n^{(h)}|~=~c_{n,h}~=~ 
{n \choose {n-h\over 2}}-{n \choose {n-h\over 2}-1}}
where $c_{n,h}$ are some generalized Catalan numbers.
(Note again that $A_n^{(h)}$ is only defined if $n=h$ mod 2.).
\par
Like in the meander case, given two arbitrary open arch 
configurations $a,b\in A_n^{(h)}$, we may consider their 
superposition (after reflecting $b$ w.r.t. the river) 
obtained by gluing their half-bridges, and connecting 
the upper and lower open arches starting from
the rightmost one so as to form $h$ arches encircling the source of 
the river.   This leads in general to a multi-component semi-meander 
formed of possibly many non-intersecting roads crossing the river, 
and possibly winding around its source, with a {\it total} 
winding number $h$. By analogy with the meander case, 
we still denote by $\kappa(a|b)$ the resulting number 
of connected components.
\par
\subsec{Meander and semi-meander determinants}
With the above definitions, let us introduce, for any given complex 
number $q$, the meander and 
semi-meander matrices ${\cal G}_{2n}(q)$ and ${\cal G}_n^{(h)}(q)$ 
of respective sizes $c_n \times c_n$ and 
$c_{n,h}\times c_{n,h}$, with entries
\eqn\mesemat{\eqalign{
\big[{\cal G}_{2n}(q)\big]_{a,b}~&=~ 
q^{\kappa(a|b)} \qquad a,b\in A_{2n} \cr 
\big[{\cal G}_n^{(h)}(q)\big]_{a,b}~&=~ q^{\kappa(a|b)} \qquad 
a,b \in A_n^{(h)} \cr}}
As $A_{2n}^{(0)}=A_{2n}$, we have 
${\cal G}_{2n}^{(0)}(q)={\cal G}_{2n}(q)$, hence 
the meander matrix is just a particular 
case of semi-meander matrix with $h=0$ 
winding number. 
Nevertheless, for a clearer exposition, we will distinguish 
between the 
two cases.
\par
Let us denote by $U_m(q)$ the Chebishev polynomials of the first kind, 
namely such that 
\eqn\cebi{U_m(2 \cos \theta)~=~{\sin(m+1)\theta \over \sin \theta}} 
for all $m \geq 0$. 
With this definition, and the integer
numbers $c_{n,m}$ defined in \oparnum,
we have the following compact formulas for the determinants of 
the matrices ${\cal G}_{2n}(q)$ and ${\cal G}_n^{(h)}(q)$
\medskip
\deb
\noindent{\bf THEOREM 1:} 
\eqn\thone{\eqalign{ 
\det \big[ {\cal G}_{2n}(q) \big]~ &=~ 
\prod_{m=1}^n \big[U_m(q)\big]^{a_{2n,2m}} \cr
a_{2n,2m}~&=~
c_{2n,2m}-c_{2n,2m+2} \cr}}
\fin
\deb
\noindent{\bf THEOREM 2:}
\eqn\thtwo{\eqalign{ \det \big[ {\cal G}_n^{(h)}(q)\big]~&=~ 
\prod_{m=1}^{{n-h \over 2}+1} \big[U_m(q)\big]^{a_{n,m}^{(h)}} \cr 
a_{n,m}^{(h)}~&=~c_{n,2m+h}-c_{n,2m+2+h} +h( c_{n,2m+h-2}-
c_{n,2m+h})\cr}}
\fin
\par 
The theorem 1 was proved in \DGGB, whereas the formula \thtwo\ 
was only conjectured there (in a slightly
different, but equivalent form).  
In the following, for pedagogical
reasons, we will first give a simplified
proof of the theorem 1, in the same spirit as \DGGB. We will 
then show how to 
generalize this proof to that of the theorem 2. Clearly, the 
theorem 2 contains 
the theorem 1 as the particular case $h=0$.
Before turning to the proofs of theorems 1 and 2 above,
we wish to provide the reader with an alternative picture
for (open or closed) arch configurations, which will prove
useful in the following. The idea is to view an (open or closed)
arch configuration of order $n$ as a walk of $n$ 
steps on a half-line.
\par
\subsec{Arch configurations and closed walk diagrams} 
There is a bijection between the arch configurations of order $2n$ 
and the closed paths of $2n$ steps on a half-line, or rather 
their two-dimensional extent, which we call a walk diagram of $2n$ 
steps. The mapping goes as follows. Let us index by $i$ the portion 
of river inbetween two consecutive bridges $i$ and $i+1$, 
$1\leq i\leq 2n-1$, by $0$ the portion to the left of the first 
bridge, and by $2n$ the
portion to the right of the last bridge.
\fig{A walk diagram of $18$ steps, and the corresponding arch 
configuration. Each dot corresponds to a segment of river. 
The height on
the walk diagram is given by the number of arches intersected by 
the vertical dotted line.}{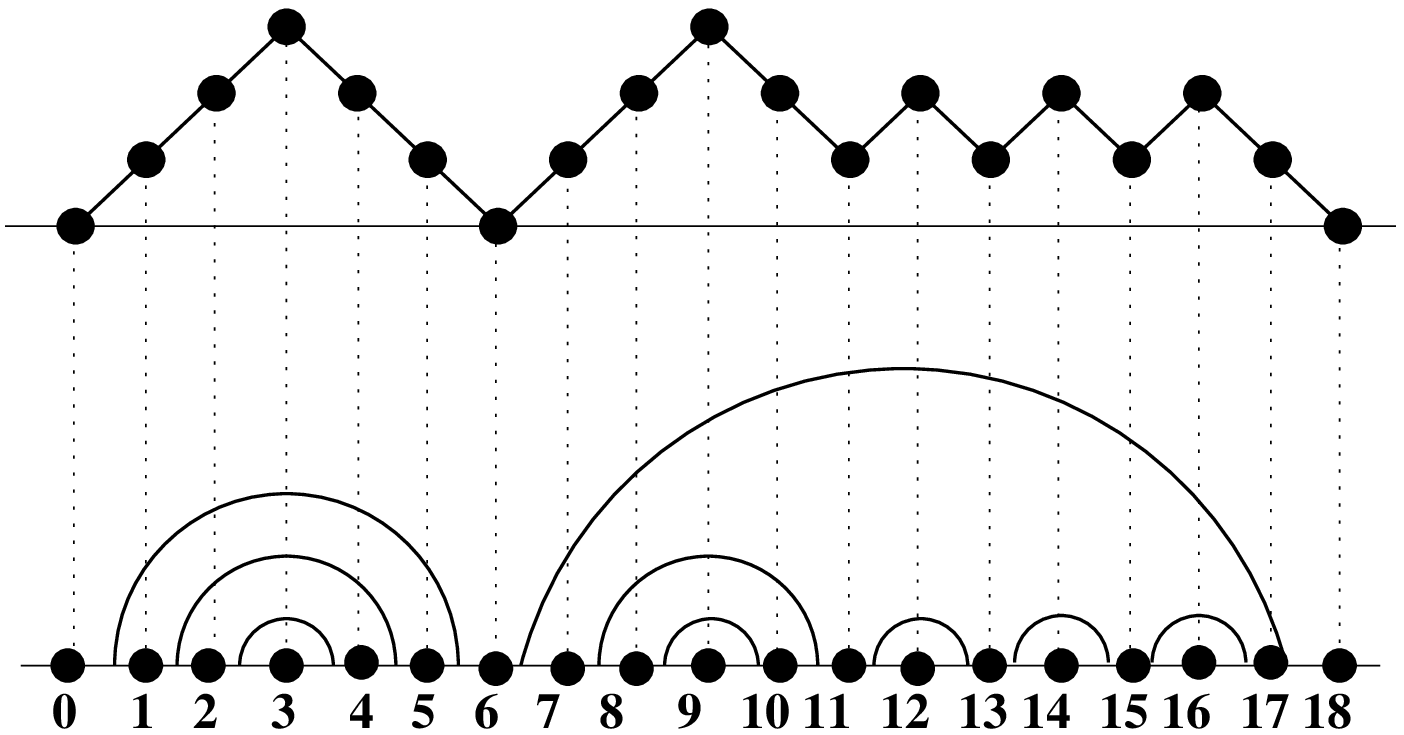}{9.cm}
\figlabel\archwalk
To each of these we associate the {\it height} $h(i)$ equal to 
the number of arches passing at the vertical of the corresponding 
portion of river.
With this definition, we have $h(0)=h(2n)=0$, $h(i)-h(i-1)=\pm 1$, 
according to whether an arch originates or terminates at the bridge 
$i$, and $h(i)\geq 0$ for all $i$. The function
$h(i)$ can be thought of as the coordinate of a walker on the 
half-line after $i$ steps. The two-dimensional extent of the 
trajectory is simply obtained by joining the consecutive points 
$(i,h(i))$ i.e., by plotting the graph of the function $h$, as 
illustrated in Fig.\archwalk.
\par
We denote by $W_{2n}$ the set of such walk diagrams of $2n$ steps, 
with $h(0)=h(2n)=0$.  In particular, the bijection implies that
$|W_{2n}|=|A_{2n}|=c_n$ for all $n\geq 0$.
In the following, we will denote indifferently 
by the same letter $a\in A_{2n}$ or $W_{2n}$ an arch configuration
or the corresponding (closed) walk diagram.
\par
\subsec{Open arch configurations and open walk diagrams}
For all $h\leq n$, $h=n$ mod 2,
there is a bijection between the set $A_n^{(h)}$ of open
arch configurations of order $n$ with $h$ open arches and
the set of open walk diagrams on a half-line,
starting at the origin and ending at height $h$ after $n$ steps.
\fig{An open walk diagram of $n=14$ steps with final height $h=4$,
and the corresponding open arch configuration.
The height on the walk diagram is given by the number of 
arches intersected by the dotted lines, {\it plus} that
of open arches lying on the left of the point 
considered.}{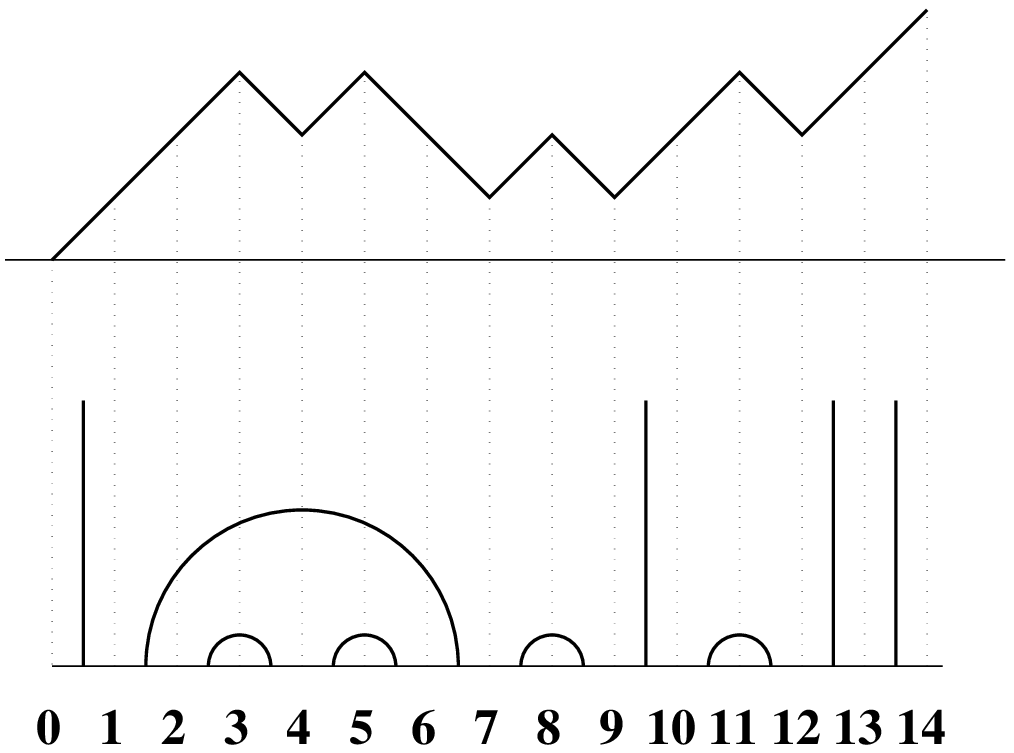}{7.cm}
\figlabel\walkop
Starting from some open arch configuration $a\in A_n^{(h)}$,
let us label as before by $0$, $1$, ..., $n$ the portions of river 
inbetween consecutive bridges of $a$ (including that to the left
of the first bridge, $0$ and to the right of the last bridge, $n$).
To each of these, we associate the {\it height} $h(i)$ equal to
the number of arches passing at the vertical of the corresponding
portion of river, {\it plus} the total number of open arches
originating from the bridges number $1$, $2$, ..., $i$, namely the
total number of open arches lying to the left of the portion $i$ of
river. With this definition, the function $i\to h(i)$ satisfies
\eqn\sath{ h(0)~=~0\ , \quad h(n)=h\ , \quad h(i)\geq 0 \ ,\quad 
{\rm and}\  h(i+1)-h(i)~=~\pm 1}
according to whether an (open or closed) arch originates from the
bridge $i$ or a (closed) arch terminates at the bridge $i$.
The function $i\to h(i)$, satisfying the properties \sath, 
defines a unique walk on the half-line, starting at the
origin (height $h(0)=0$), and ending after $n$ steps of $\pm 1$
in height at height $h(n)=h$.  The graph of the function,
$(i,h(i))$, with consecutive 
points linked by segments of line, is the two-dimensional
extent of such a walk, which we call an
{\it open walk diagram} of $n$ steps
with final height $h$.
\par
We denote by $W_n^{(h)}$ the set of open walk diagrams of $n$ steps
with final height $h$ (note that this is only defined for $h=n$ mod 2).
In particular, the above bijection implies 
\eqn\bijwa{ |W_n^{(h)}|~=~|A_n^{(h)}|~=~ c_{n,h} }
In the following, we 
will also use indifferently the same letter $a$ to denote
an element of $W_n^{(h)}$ or $A_n^{(h)}$ whichever picture
is most convenient.
\par
\newsec{The meander determinant: proof of theorem 1}
In this section, we give a detailed proof of theorem 1. We first 
recall the equivalence between arch configurations and reduced 
elements of the Temperley-Lieb algebra $TL_n(q)$, or rather
a certain left ideal ${\cal I}_n(q)$ of $TL_{2n}(q)$, isomorphic
to $TL_n(q)$. In the latter 
language, the meander matrix ${\cal G}_{2n}(q)$ \mesemat\ is interpreted
as the Gram matrix of a basis (called basis 1) of ${\cal I}_n(q)$
with respect to the standard bilinear form. The determinant of 
${\cal G}_{2n}(q)$ will be a by-product of the orthogonalization 
of this matrix. 
\par
\subsec{Temperley-Lieb algebra and arch configurations}
The arch configurations of order $2n$ have a direct interpretation 
in terms of reduced elements of the Temperley-Lieb algebra
$TL_n(q)$, for a given complex number $q$.
The latter is best expressed in its pictorial form, as acting on
a ``comb" of $n$ strings, with the $n$ 
generators $1$, $e_1$, $e_2$, ..., $e_{n-1}$ defined as
\eqn\braid{ 1~=~\figbox{2.cm}{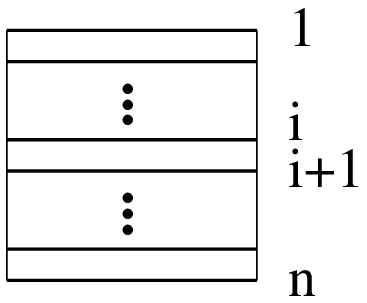} 
\qquad e_i~=~\figbox{2.cm}{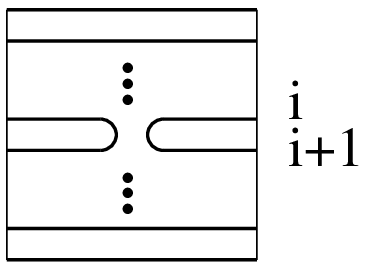} }
The most general element $e$ of $TL_n(q)$ is obtained by 
composing the generators \braid\ like dominos. 
The algebra is defined through the following 
relations between the generators
\eqn\tla{\eqalign{(i)\ \ \ \ \ \ \ \ \  e_i^2 ~&=~ q \, e_i 
\quad i=1,2,...,n-1\cr
(ii)\ \ \ \ [e_i,e_j]~&=~0 \quad {\rm if}\ |i-j|>1 \cr
(iii)\ e_i\, e_{i \pm 1}\, e_i~&=~ e_i  \quad i=1,2,...,n-1\cr}} 
The relation (ii) expresses the locality of the $e$'s,
namely that the $e$'s commute whenever they involve distant strings. 
The relations (i) and (iii) read respectively 
\eqn\unbraid{\eqalign{(i)\ \ \ \ \ \ \ \ \ e_i^2~&=~ 
\figbox{2cm}{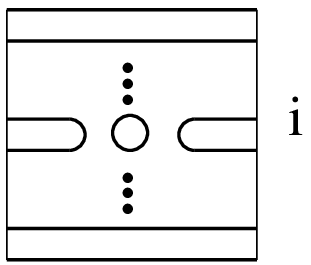}~=~q~\figbox{2.4cm}{ei.eps}~=~q 
\, e_i\cr (iii)\ e_i\, e_{i+1}\, e_i~&=~ 
\figbox{2.4cm}{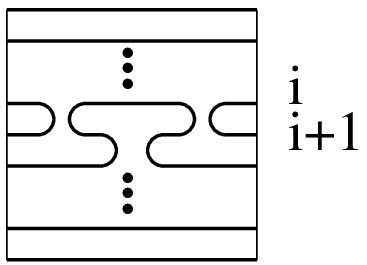}~=~\figbox{2.4cm}{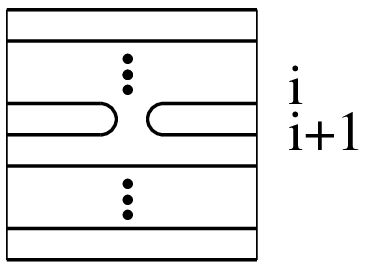}~=~ e_i\cr}}
In (i), we have replaced a closed loop by a factor $q$.  Therefore
we can think of $q$ as being a weight per connected component 
of string. In (iii), we have simply ``pulled the string" number $i+2$.
\par
An element $e\in TL_n(q)$ is said to be reduced if all its 
strings have been pulled and all its loops removed, and if 
it is further normalized so as to read $\prod_{i\in I} e_i$ 
for some minimal
finite set of indices $I$. A reduced element is formed of 
exactly $n$ strings.
\fig{The transformation of a reduced element of $TL_9(q)$
into an arch configuration of order $18$. The reduced
element reads $e_3 e_4 e_2 e_5 e_3 e_1 e_6 e_4 
e_2$.}{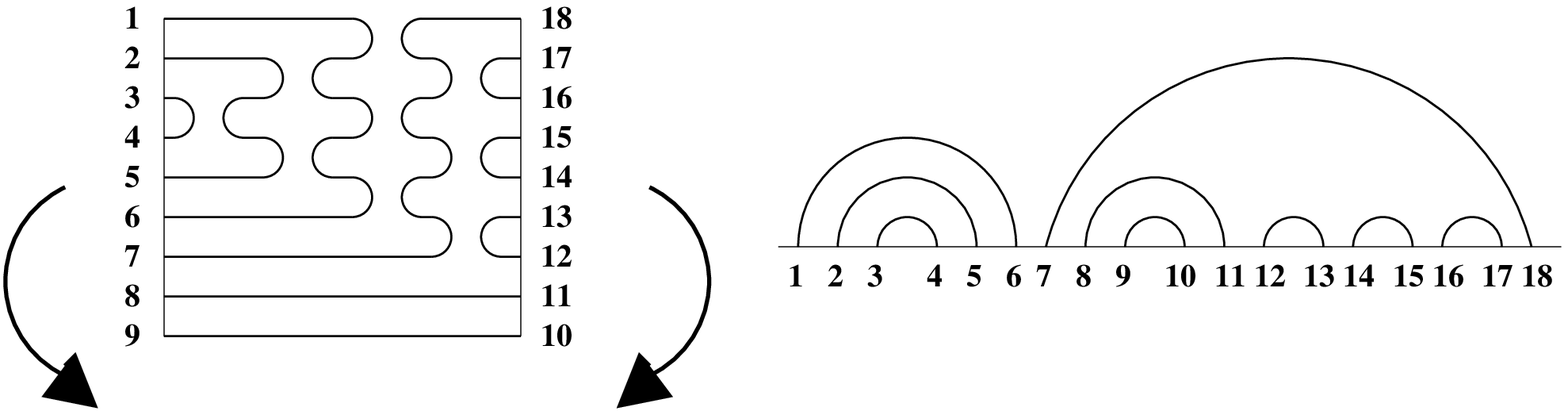}{10.cm} \figlabel\artla
There is a bijection between the reduced elements of 
the Temperley-Lieb algebra
$TL_n(q)$ and the arch configurations of order $2n$.
Starting from a reduced element of $TL_n(q)$, we index the 
left ends of the $n$ strings by $1,2,...,n$, and the right 
ends of the strings 
$2n,2n-1,...,n+1$ from top to bottom (see Fig.\artla\ for an 
illustration). Interpreting these ends as bridges, and 
aligning them on a line, 
we obtain a planar pairing of
bridges by means of non-intersecting strings (arches), 
hence an arch configuration of order $2n$.
Conversely, we can deform the arches of any arch configuration 
of order $2n$ to form a reduced element of $TL_n(q)$.
As a consequence, we have dim$(TL_n(q))=c_n$, as vector space 
with a basis formed by all the reduced elements.
\par
In the following, we will rather use the identification 
between $TL_n(q)$ and the left ideal ${\cal I}_n(q)$ of 
$TL_{2n}(q)$ generated by the element $u_n=e_1 e_3 ... e_{2n-1}$, 
which goes over to reduced elements. 
\par
\fig{The arch configuration of order $18$ of Fig.\artla\
is immediately interpreted as an element of the ideal 
${\cal I}_9(q)$ of $TL_{18}(q)$, by adding a succession 
of $n=9$ strings linking the consecutive upper ends of 
strings by pairs (for simplicity, the element of $TL_{18}(q)$
is now read from bottom to top). 
The corresponding reduced element of 
${\cal I}_9(q)$ reads, from bottom to top,
$(e_3 e_9)(e_2 e_4 e_8 e_{10}e_{12} e_{14} e_{16}) 
(e_1 e_3 e_5 e_7 e_9 e_{11} e_{13} e_{17})$. }{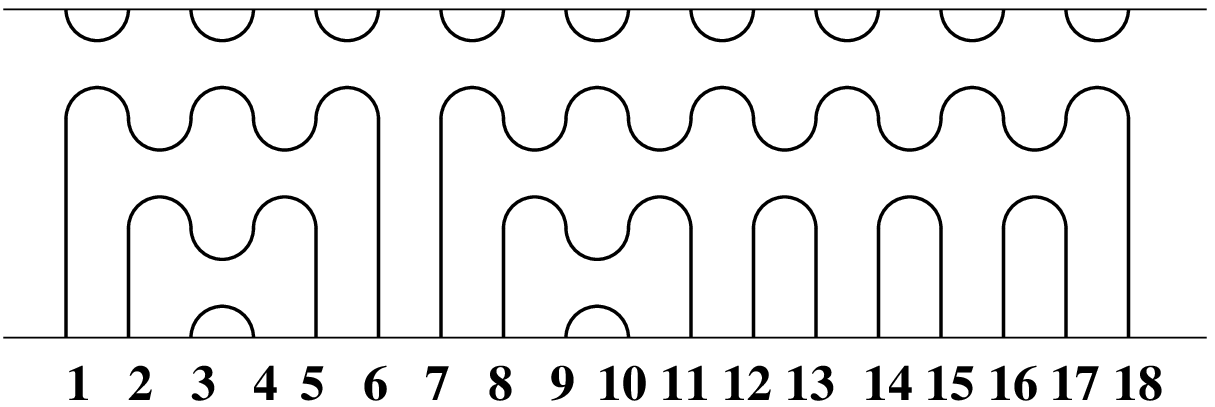}{7.cm} 
\figlabel\ideal
Indeed any reduced element of ${\cal I}_n(q)$ has a pictorial 
representation as a set of $2n$ strings linking by pairs the 
$2n$ left and $2n$ right ends of strings, as illustrated in 
Fig.\ideal. The pairing of the righ ends of strings is very 
simple, and represents the right factor $e_1 e_3...e_{2n-1}$. 
It consists of $n$ arches connecting the $n$ pairs of 
successive right ends of strings. Therefore the $2n$ 
left ends of strings are connected among themselves 
through the $n$ remaining strings. This gives exactly 
an arch configuration of order $2n$. 
\par
\fig{Example of a walk diagram in $W_{8}$, expressed as the 
result of four box additions on the fundamental walk 
$a_4$.}{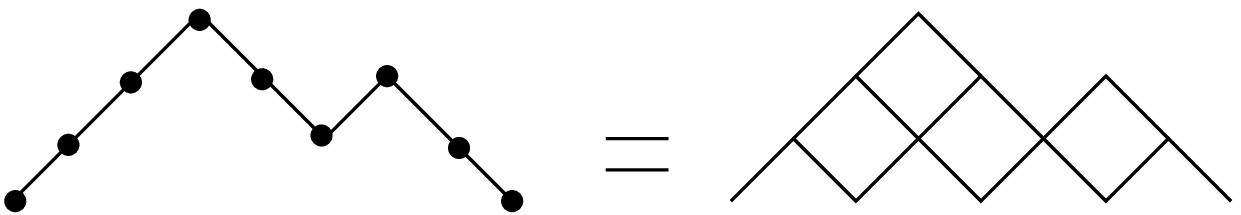}{8.cm}
\figlabel\boxad
The converse construction is best expressed in the walk diagram 
representation of arch configurations.
Let us construct a map $\rho$ from $W_{2n}$
to the set of reduced elements of the ideal ${\cal I}_n(q)$
of $TL_{2n}(q)$.
Let $a_n$ denote the {\it fundamental} walk diagram of $W_{2n}$, 
such that 
\eqn\fundam{a_n\ :\ h(0)=h(2)=...=h(2n)=0 \quad {\rm and} \quad
h(1)=h(3)=...=h(2n-1)=1} 
To this diagram, we associate the reduced element 
\eqn\mapo{u_n~=~\rho(a_n)~=~e_1e_3...e_{2n-1}}
Now any walk diagram in $W_{2n}$ is obtained from $a_n$ by 
successive {\it box additions}, illustrated in Fig.\boxad. 
A box addition at a 
minimum $i$ of $a \in W_{2n}$, i.e. where $h(i+1)=h(i-1)=h(i)+1$, 
simply consists in shifting $h(i)\to h(i)+2$, which amounts 
to formally add a 
square box which fills the minimum at $i$ and transforms 
it into a maximum. We will denote by $a \to a+\diamond_i$ this 
operation on $a\in W_{2n}$ (this notation keeps
track of the point $i$ at the vertical of which the box is added).
This enables us to define the {\it length} of a walk
diagram $a \in W_{2n}$ as the number of box additions which have
to be performed on the fundamental $a_n$ to build $a$. We set
\eqn\lengta{ |a|~=~ \#{\rm boxes}\ {\rm in}\ a \qquad {\rm for}\
{\rm all}\ a \in W_{2n} }
In particular, we have $|a+\diamond_i|=|a|+1$.
The mapping $\rho$ is then defined as 
\eqn\mapro{\rho(a+\diamond_i)~=~ e_i \ \rho(a) }
where the box addition is made at the point $i$ (necessarily a 
minimum of $a$). 
\fig{Example of the mapping $\mu$ between an element
$a\in W_{8}$ and $e=\rho(a)\in{\cal I}_4(q)$.
We read the element $e=\rho(a)$ from the various layers of 
box additions, using the formula \mapro. Here we get 
$e=(e_3)(e_2 e_4 e_6)(e_1 e_3 e_5 e_7)$ (the parentheses correspond
to the successive layers of boxes added.}{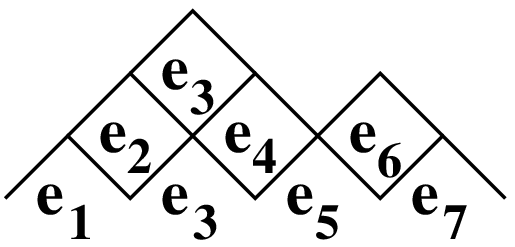}{5.cm} 
\figlabel\boxadideal
The most general reduced element $e=\rho(a)$ in ${\cal I}_n(q)$, 
$a\in W_{2n}$, is therefore
written as the product over all box additions 
leading from the fundamental walk $a_n$ to $a$, of the corresponding 
$e_i$'s. This is illustrated in Fig.\boxadideal.
The reduced elements of ${\cal I}_n(q)$ form a basis (which we 
call {basis 1} from now on) of the corresponding vector space 
over the complex numbers.
We have established that $|W_{2n}|={\rm dim}({\cal I}_n(q))=c_n$. 
For simplicity, we will adopt the following notation for the 
basis 1 elements: we write
\eqn\notat{ (a)_1~=~ \rho(a) \qquad {\rm for}\ {\rm any} \ a\in W_{2n}}
As an example, the basis 1 for ${\cal I}_3(q)$ is formed by the
$c_3=5$ following elements
\eqn\elembato{\eqalign{
\left( \figbox{2.cm}{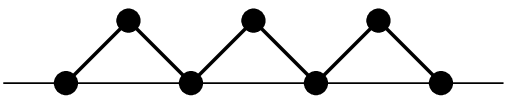} \right)_1~&=~ e_1 e_3 e_5 \cr
\left( \figbox{2.cm}{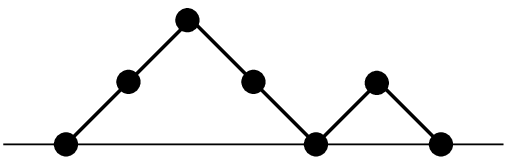} \right)_1~&=~ e_2 e_1 e_3 e_5 \cr
\left( \figbox{2.cm}{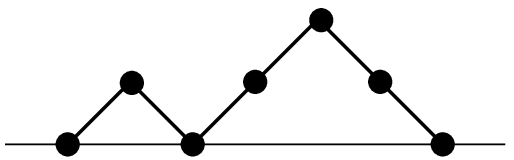} \right)_1~&=~ e_4 e_1 e_3 e_5 \cr
\left( \figbox{2.cm}{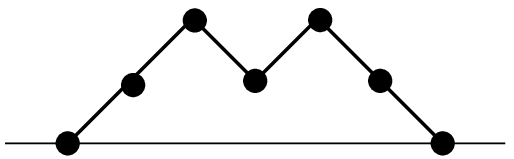} \right)_1~&=~ e_2 e_4 e_1 e_3 e_5 \cr
\left( \figbox{2.cm}{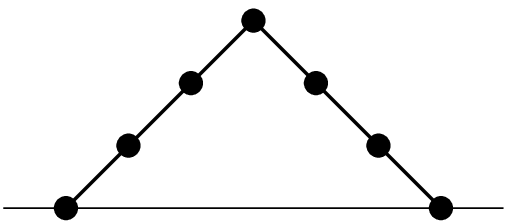} \right)_1~&=~ e_3 e_2 e_4 e_1 e_3 e_5 \cr}}
indexed by the $5$ walk diagrams of $W_6$.
\par
We now show how to reconstruct the string-domino pictorial representation
attached to an element of ${\cal I}_{n}(q)$, from the box decomposition
of the corresponding walk $a\in W_{2n}$.
The idea is to represent the $e_i$'s
forming the fundamental element $u_n$ \mapo\ 
by boxes as well. Actually each box
will have the meaning of a left multiplication by $e_i$, this time
acting on $1$.
Starting from some walk diagram $a\in W_{2n}$, we write it as the result
of box additions on the empty diagram. 
To go to the string-domino picture, we have to draw ``arches" using
the box configurations. This is done by marking each box 
with a pair of strings as follows
\eqn\symbox{ \figbox{1.cm}{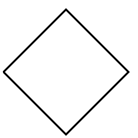} ~\to ~ \figbox{1.cm}{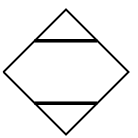}}
and by continuing each string with a vertical line ending at some
string-end on the border of the corresponding domino.
\fig{The box decomposition of an element $a\in {\cal I}_{5}(q)$,
and the corresponding string-domino picture.
Note that the domino is now read from top to bottom,
rather than from bottom to top as it used to be in 
Fig.\ideal.}{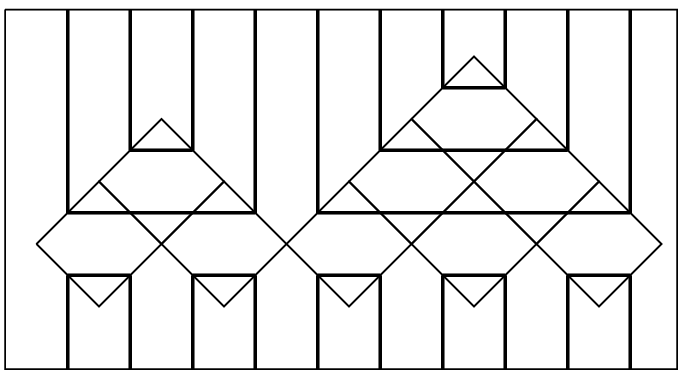}{6.cm}
\figlabel\strido
This is illustrated in Fig.\strido, where the strings are represented in
thick black lines.
\par
\subsec{Gram matrix for the basis 1 of ${\cal I}_n(q)$} 
\fig{Computing the trace of a reduced element $e$ of $TL_6(q)$: 
(i)connect the left and right ends of strings (dashed lines)
(ii) count the number of connected components of string: we
find $\kappa(e)=3$ here. This leads to the trace
${\rm Tr}(e)=q^3$.}{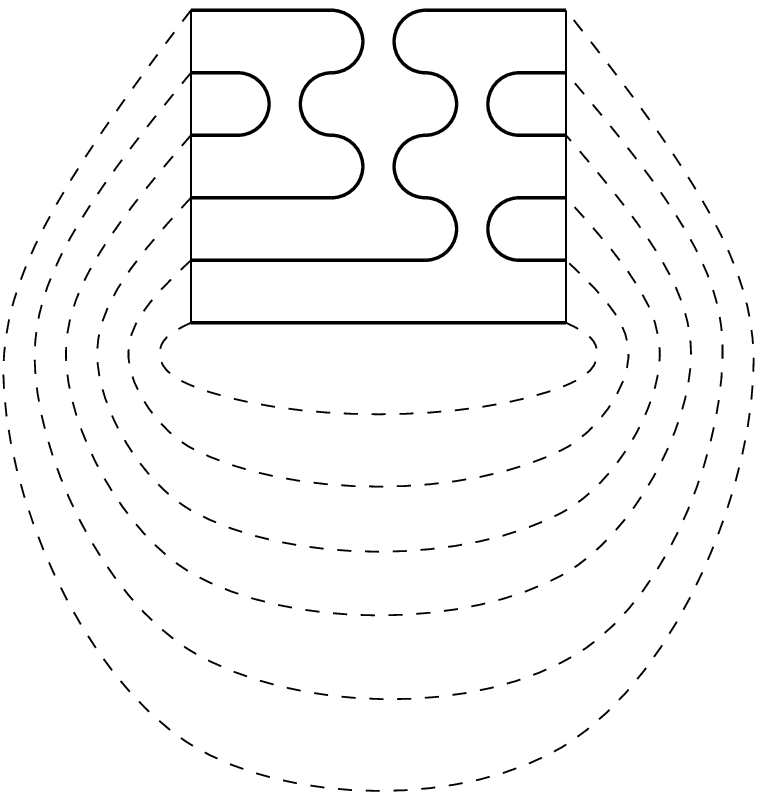}{5.cm}
\figlabel\trace
The Temperley-Lieb algebra is endowed with a 
standard trace, defined on the reduced elements as the number 
\eqn\tradef{ {\rm Tr} (e)~=~ q^{\kappa(e)} }
where $\kappa(e)$ is the number of connected components of strings 
after the identification of the left ends of strings with the right 
ones, as depicted in Fig.\trace. This definition extends by 
linearity to any element of $TL_n(q)$. 
Given a reduced element $e\in TL_n(q)$, we may consider the adjoint 
$e^t$, obtained by reflecting the corresponding
arch configuration w.r.t. the river. The corresponding operation
on $TL_n(q)$ satisfies $e_i^t=e_i$ (the generators are self-adjoint), 
and $(ef)^t=f^t e^t$ for any reduced elements $e,f$. 
Taking the adjoint simply reflects the string-domino picture of the 
corresponding reduced element, and exchanges the left and right 
ends of strings. 
Again, this extends
to any element of $TL_n(q)$ by linearity. 
We can now introduce the bilinear form
\eqn\bilin{ (e,f)~=~ {\rm Tr} 
(e f^t) \qquad {\rm for} \ {\rm any} \ e,f \in TL_n(q)}
\par
The 
above definitions extend by restriction to any ideal of the 
Temperley-Lieb algebra. Let us now concentrate on the ideal 
${\cal I}_n(q)$ of $TL_{2n}(q)$.
Let us consider the Gram matrix of the basis 1, with respect 
to the bilinear form \bilin, namely the matrix $\Gamma_{2n}(q)$, 
with entries 
\eqn\gramean{ \big[ \Gamma_{2n}(q) \big]_{a,b}~=~ 
\big( (a)_1,(b)_1\big) ~=~ {\rm Tr}\big( (a)_1 (b)_1^t \big) } 
where $a,b$ run over the walk diagrams of $W_{2n}$ which are 
used to index the corresponding basis 1 elements $(a)_1,(b)_1$. 
\par
\fig{The bilinear form $(e,f)$ is obtained by first multiplying 
$e$ with $f^t$, and then identifying the upper and lower ends of 
the strings (The bridges numbered $1,2,...,10$ are identified), 
and counting the number of 
connected components of strings. Here we have created $n=5$
simple loops at the connection between the two
dominos, and $\kappa(a|b)=3$ other loops, from the
superposition of the arch configurations 
$a$ and $b$ of order $10$, corresponding respectively to $e$
and $f$. 
Note that $b\to b^t$ is reflected 
w.r.t. the river, corresponding to the adjoint in 
$f^t$.
Finally we have ${\rm Tr}(e f^t)=q^{n+\kappa(a|b)}=q^8$ here.
}{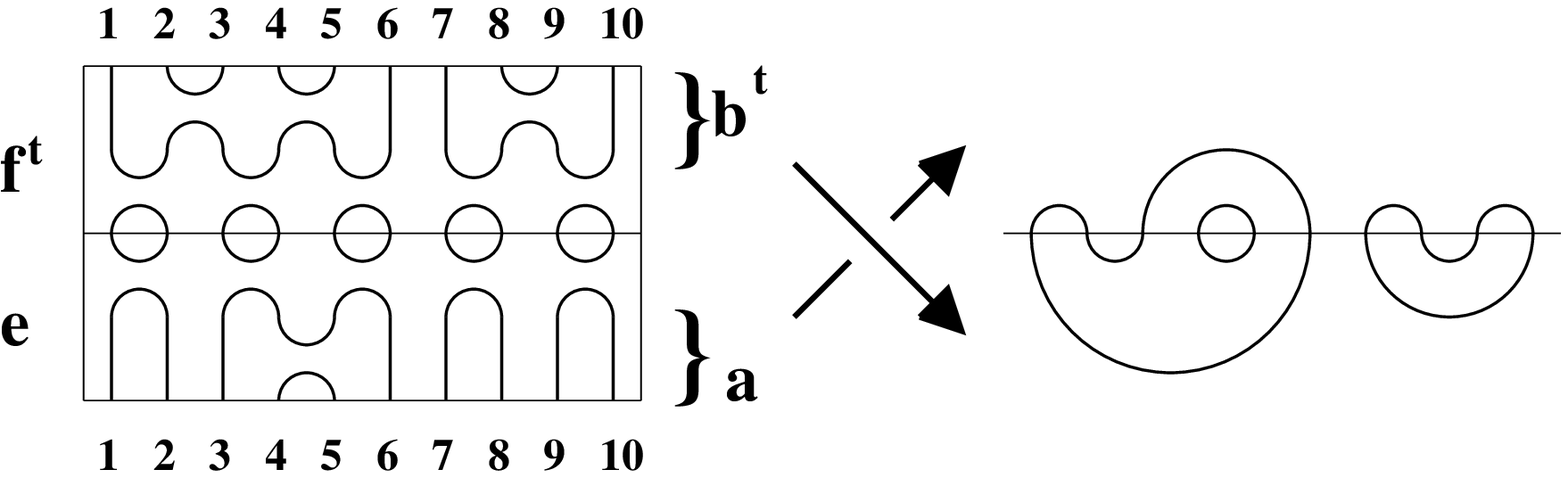}{10.cm} 
\figlabel\evatra
But evaluating the matrix element \gramean\
just amounts, as illustrated in Fig.\evatra, 
to connecting the 
domino corresponding to $(a)_1$ and the reflection of the domino 
corresponding to $(b)_1$, and identifying the left 
ends of strings (on the bottom of the figure) 
with the right ones (on top of the figure), and counting the number of 
connected components of strings.
Because of the particular form of the elements of ${\cal I}_n(q)$ 
(see Fig.\ideal), this procedure will create $n$ loops
along the connection line between the two dominos 
(these loops
are formed by the arches connecting consecutive ends of strings) 
plus an extra
$\kappa(a|b)$ loops, namely those appearing in the 
superposition of the arch configurations $a$ and 
(the reflection of) $b$. Therefore we have
\eqn\gamg{\big[ \Gamma_{2n}(q)\big]_{a,b}~=~ 
q^{n+\kappa(a|b)} ~=~ q^n \big[{\cal G}_{2n}(q)\big]_{a,b}} 
by comparison with the previous definition \mesemat\ of 
the meander matrix ${\cal G}_{2n}(q)$.
\par
Hence the meander determinant is simply related to 
the Gram determinant of the basis 1, through
\eqn\detgra{ \det(\Gamma_{2n}(q))~=~ q^{n c_n} \det{\cal G}_{2n}(q) }
The remaining subsections of this section will be devoted to the
computation of the Gram determinant of the basis 1 of ${\cal I}_n(q)$.
\par
\subsec{Orthonormalization of the basis 1}
In the following,
we will compute the Gram determinant \detgra\ by performing an
explicit Gram-Schmidt
orthonormalization of the basis 1 w.r.t. the bilinear form
\bilin.
The orthonormalization process consists in a change of basis
from the basis 1 to another basis, which we call basis 2, satisfying
the following properties
\item{(i)} The basis 2 elements are still indexed by the
walk diagrams of $W_{2n}$, we denote them by $(a)_2$, $a \in W_{2n}$.
\item{(ii)} The basis 2 is orthogonal w.r.t. the bilinear form \bilin,
namely $\big( (a)_2, (b)_2\big)=0$ whenever $a\neq b$.
\item{(iii)} The basis 2 elements have all the same norm $1$, namely
\eqn\nortwo{ \big( (a)_2,(a)_2\big)~=~ 1 \qquad {\rm for}\ {\rm any}\
a\in W_{2n}}
\par
The basis 2 elements are constructed as follows. We start from the 
{\it fundamental} element $(a_n)_2$, indexed by the fundamental
walk of $W_{2n}$ \fundam, and defined as
\eqn\fundto{ (a_n)_2~=~ q^{-n} e_1 e_3 ...e_{2n-1} }
The normalization factor ensures that the property $(iii)$ above
holds for the norm of this element, namely that
\eqn\norinit{\big( (a_n)_2,(a_n)_2\big)~=~1} 
(Indeed, $(a_n)_2 (a_n)_2^t=(a_n)_2$,
and ${\rm Tr}(a_n)_2=q^{-n} q^n=1$.).
As any walk diagram $a\in W_{2n}$ is obtained from the fundamental one $a_n$
by successive box additions, we define the other basis 2 elements
by the following box addition rule, which amounts to a recursion. 
Suppose we have constructed $(a)_2$ for some $a \in W_{2n}$.
The following rule gives the element $(a+\diamond_{i,\ell})_2$, where
a box addition has been performed on a minimum $i$ of $a$,
with $h(i+1)=h(i-1)=h(i)+1=\ell$, the {\it height} of the box
addition.
\eqn\boxrule{ (a+\diamond_{i,\ell})_2~=~ \sqrt{\mu_{\ell+1}\over
\mu_{\ell}} \big( e_i -\mu_{\ell} \big)\ (a)_2 }
where we have used the notation
\eqn\notmu{ \mu_\ell~=~ {U_{\ell-1}(q) \over U_{\ell}(q)} \qquad
{\rm for}\ \ell=1,2,3...}
in terms of the Chebishev polynomials \cebi.  
Note that the recursion relation $U_{m+1}(q)=q U_m(q)-U_{m-1}(q)$
translates into the relation
\eqn\murela{ {1 \over \mu_1}-\mu_m ~
=~{1 \over \mu_{m+1}}\qquad {\rm for} \ {\rm all}\ m\geq 1}
The rule \boxrule\
may be viewed as a deformation of the rule \mapro\ used to construct
the basis 1. However, two new ingredients have appeared: 
(i) the
box addition now depends on the height $\ell$
at which it is performed (hence the notation $\diamond_{i,\ell}$,
to keep track of this height) and (ii) there is an overall change
of normalization $\sqrt{\mu_{\ell+1}/\mu_\ell}$.
Together with the initial point \fundto, the recursive rule \boxrule\
determines the basis 2 elements completely. 
\par
By construction,
these elements all have the right factor $e_1 e_3...e_{2n-1}$,
hence belong to the ideal ${\cal I}_n(q)$. Moreover, when expressed
on basis 1 elements, they read
\eqn\chbas{ (a)_2~=~\sum_{b \subset a\atop
b\in W_{2n}} P_{b,a} (b)_1 }
where the matrix elements of $P$ are products of factors involving 
the $\mu_m$'s, and the sum extends only on the walk diagrams $b$
included in $a$, i.e., such that $a$ is obtained from $b$ by some 
box additions (this includes the case $b=a$ with no box addition).
The change of basis 1 $\to$ 2 is therefore triangular, as 
the walk diagrams $a\in W_{2n}$
may be arranged by growing length $|a|$ \lengta, 
thus making the matrix $P$ of the change of basis upper triangular
(this
implies that the basis 2 is indeed a basis of ${\cal I}_n(q)$).
\par
To distinguish between the two different box additions
\mapro\ for basis 1 and \boxrule\ for basis 2, which could
both be performed on a given walk diagram $a \in W_{2n}$, representing
either a basis 1 or a basis 2 element,
we decide to represent \mapro\ by grey boxes, whereas
\boxrule\ is represented by white boxes, namely 
\eqn\notboru{ e_i ~=~ \figbox{1.cm}{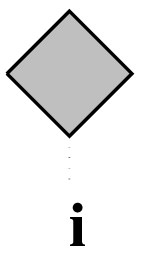} \quad {\rm and} \quad
\sqrt{\mu_{m+1}\over \mu_m} \big( e_i -\mu_m\big)~=~
\figbox{1.7cm}{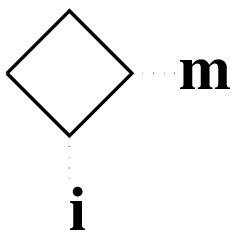} }
(We have indicated the position $i$ and the height $m$ at which the
box acts.) 
In this pictorial representation, the basis 2 elements of
${\cal I}_3(q)$ read
\eqn\batorep{\eqalign{
\left( \figbox{1.5cm}{e1w.eps} \right)_2~&=~\mu_1^3 
\figbox{1.5cm}{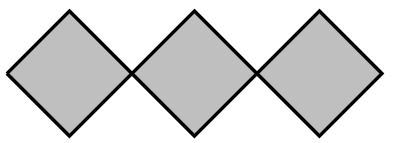}~=~
\mu_1^3 e_1 e_3 e_5 \cr
\left( \figbox{1.5cm}{e21w.eps} \right)_2~&=~\mu_1^3 
\figbox{1.5cm}{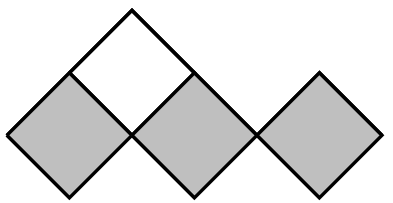}~=~
\mu_1^{5/2}\mu_2^{1/2} (e_2-\mu_1) e_1 e_3 e_5 \cr
\left( \figbox{1.5cm}{e12w.eps} \right)_2~&=~\mu_1^3 
\figbox{1.5cm}{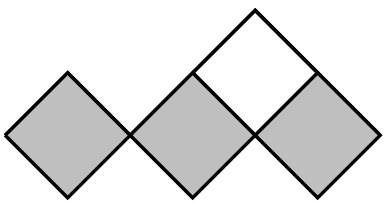}~=~
\mu_1^{5/2}\mu_2^{1/2} (e_4-\mu_1) e_1 e_3 e_5 \cr
\left( \figbox{1.5cm}{e2w.eps} \right)_2~&=~\mu_1^3 
\figbox{1.5cm}{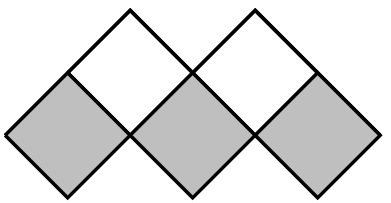}~=~
\mu_1^{2}\mu_2 (e_2-\mu_1) (e_4-\mu_1) e_1 e_3 e_5 \cr
\left( \figbox{1.5cm}{onew.eps} \right)_2~&=~\mu_1^3 
\figbox{1.5cm}{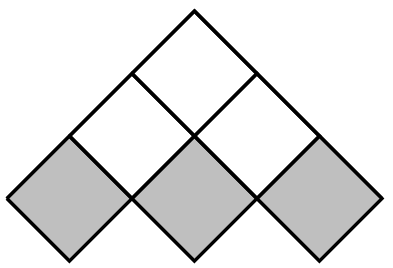}~=~
\mu_1^{2}\mu_2^{1/2} \mu_3^{1/2} 
(e_3-\mu_2)(e_2-\mu_1)(e_4-\mu_1) e_1 e_3 e_5 \cr}}
\par
With these definitions, we have the
\par
\deb
\noindent{\bf PROPOSITION 1:}
\par
The basis 2 is orthonormal w.r.t. the bilinear
form \bilin, namely 
\eqn\bilorto{ \big( (a)_2,(b)_2\big)~=~ \delta_{a,b} \qquad
{\rm for}\ {\rm all}\ a,b \in W_{2n}}
\fin
\par
We will first prove by recursion, using box additions,
the following
\par
\deb
\noindent{\bf LEMMA 1:}
\eqn\leone{ (a)_2^t(b)_2~=~0 \qquad {\rm for}\ {\rm all}\ 
a, b \in W_{2n}\ {\rm such}\ {\rm that}\ |a|\leq|b|\ {\rm and}\  
a\neq b}
\fin
\par
Note that this result is stronger than the one for the trace
of $(a)_2^t (b)_2$, which is implied in proposition 1. We learn
from lemma 1 that the product of any two distinct basis 2 elements
vanishes. This stronger result is linked to the property that ${\cal
I}_n(q)$ is an ideal. This remark will take its full strength when
we study the semi-meander determinant.  
For an element $(a)_2$ of basis 2, the length of $a$, $|a|$, represents
its number of white boxes. We will therefore prove the lemma 1 by
recursion on the white box addition. 
\par
Suppose that the property $\cal P$
\eqn\prorec{ ({\cal P}): (a)_2^t(b)_2~=~0 \quad {\rm for}\ {\rm all}\ 
b\ {\rm such}\ {\rm that}\ |b|\geq|a|\ {\rm and}\ b\neq a }
holds for some $a\in W_{2n}$.  Let us prove that $\cal P$
holds for $a+\diamond$, for any white box addition on $a$.
Pick any walk diagram $b\in W_{2n}$ such that 
\eqn\hypolen{|b|~\geq~|a+\diamond|~=~|a|+1}
We wish to evaluate the quantity
\eqn\evapro{ (a+\diamond_{i,\ell})_2^t (b)_2 }
and show that it vanishes. The idea is simply to {\it transfer}
the box addition from $a$ to $b$, namely use the commutation of $e_i$
with $e_j$, $|j-i|>1$, to let the white box $\diamond_{i,\ell}$ 
act on $(b)_2$ 
(the white box is self-adjoint, and multiplies $(a)_2^t$ to the 
right, hence we can let it act on $(b)_2$ by left multiplication).
There are however two problems associated with this transfer:
(i) $b$ may not have a minimum at $i$ or (ii) if $b$ has
a minimum at $i$, it may lie at a different height $m \neq \ell$.
We therefore have to distinguish between the following
three possible
configurations of $b$ at $i$ (maximum, slope, minimum)
\par
\noindent{$\ \ \ \ \ \ $ 
(1)} $b$ has a {\it maximum} at $i$, 
namely with $h(i+1)=h(i-1)=h(i)-1=m$.
This means that $b$ itself is the result of a box addition at $i$
on the walk $b'=b-\diamond_{i,m}$ with $h(i+1)=h(i-1)=h(i)+1=m$, and
all other $h(j)$ identical to those of $b$. Hence the white box addition
on $(b)_2$ reads
\eqn\leftmul{\eqalign{ \sqrt{\mu_{\ell+1} \over \mu_\ell}\big( 
e_i -\mu_\ell\big) (b)_2~&=~ 
\sqrt{\mu_{\ell+1}\mu_{m+1} \over \mu_\ell \mu_m}
(e_i-\mu_\ell)(e_i-\mu_m)(b-\diamond_{i,m})_2\cr
&=~\sqrt{\mu_{\ell+1}\mu_{m+1} \over \mu_\ell \mu_m}
\bigg[(\mu_1^{-1}-\mu_\ell-\mu_m)(e_i-\mu_m)\cr
&+(\mu_1^{-1}-\mu_m)\mu_m\bigg]
(b-\diamond_{i,m})_2 \cr
&=~\sqrt{\mu_{\ell+1} \over \mu_\ell}(\mu_{m+1}^{-1}-\mu_\ell)(b)_2+
\sqrt{\mu_{\ell+1}\mu_{m} \over \mu_\ell \mu_{m+1}}
(b-\diamond_{i,m})_2\cr}}
where we have used the relation \murela, and the property
$e_i^2=q e_i=\mu_1^{-1}e_i$. In the second line of 
\leftmul, we have reconstructed a white box addition on the minimum
of $b-\diamond_{i,m}$ at $i$ (first term) up to an additive
constant (second term), resulting respectively in the terms
$(b)_2$ and $(b-\diamond_{i,m})_2$ of the result.
\par
\noindent{$\ \ \ \ \ \ $ 
(2)} $b$ has an {\it ascending slope} (resp. a {\it descending
slope}) at $i$, namely with $h(i+1)-1=h(i)=h(i-1)+1=m$ 
(resp. $h(i+1)+1=h(i)=h(i-1)-1=m$). In either case, let us write  
the white box addition on $(b)_2$ as
\eqn\witadb{ \sqrt{\mu_{\ell+1} \over \mu_\ell} e_i 
-\sqrt{\mu_\ell \mu_{\ell+1}}}
namely as a term proportional to a grey box addition 
(multiplication by $e_i$) plus a constant.  But the grey box addition 
on a white slope of $(b)_2$ has a zero result.
Indeed, the slope is itself the result of prior white box additions,
hence, in the case of an ascending slope
\eqn\greybw{\eqalign{ \figbox{1.8cm}{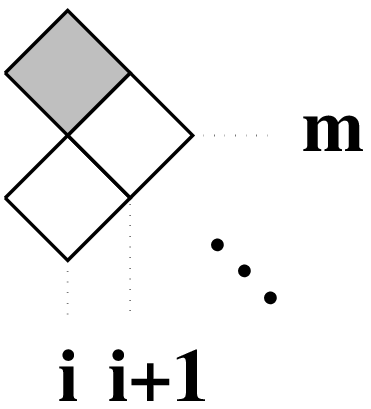}~&
=~e_i \sqrt{\mu_{m+1}\over \mu_{m-1}}
(e_{i+1}-\mu_m)(e_i-\mu_{m-1})... \cr
&=~\sqrt{\mu_{m+1}\over \mu_{m-1}}\big( (1-\mu_1^{-1}\mu_m
+\mu_m \mu_{m-1}) e_i-\mu_{m-1} e_i e_{i+1} \big)...      \cr
&=-\sqrt{\mu_{m+1}\mu_{m-1}} e_i e_{i+1}... \cr}}
where we have used the relations $e_i^2=\mu_1^{-1} e_i$ and 
$e_i e_{i+1} e_i=e_i$, and where \murela\ has implied the vanishing
of the coefficient of $e_i$ in the second line of \greybw.
We are left with an expression involving the action of a grey
box at the point $i+1$ (factor $e_{i+1}$) 
on a white slope of the rest of $(b)_2$
(symbolized by the ... in \greybw). We can therefore repeat the
calculation \greybw\ with $i\to i+1$ (and $m\to m-1$), and
so on, until the ``bottom" of the diagram is reached, namely
the situation where the slope is formed by the piling up of
a grey and a white box:
\eqn\nediagr{\eqalign{ \figbox{1.8cm}{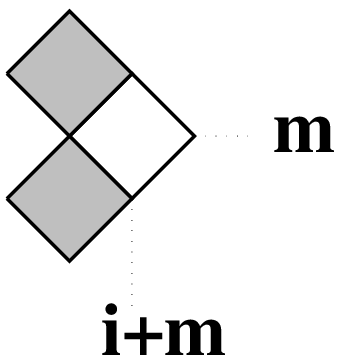}~&=~ e_{i+m-1} 
\sqrt{\mu_2 \over \mu_1}(e_{i+m}-\mu_1)e_{i+m-1}... \cr
&=~0\cr}}
by using $e_{i+m-1} e_{i+m} e_{i+m-1}=e_{i+m-1}$ and 
$e_{i+m-1}^2=\mu_1^{-1} e_{i+m-1}$. The same reasoning applies
for a descending slope: such a slope may indeed be viewed as the
adjoint of an ascending slope, whereas the white box to be added
is self-adjoint.
The addition of a white box on any slope of $b$ therefore
reduces to the second term of \witadb, namely
\eqn\adbwit{ \sqrt{\mu_{\ell+1}\over \mu_{\ell}} 
\big( e_i -\mu_\ell\big) (b)_2~=~-\sqrt{\mu_{\ell+1}\mu_\ell} (b)_2}
\par
\noindent{$\ \ \ \ \ \ $ 
(3)}
$b$ has a {\it minimum} at $i$, namely with $h(i+1)=h(i)+1=h(i-1)=m$.
Then, writing
\eqn\wrifty{
\sqrt{\mu_{\ell+1} \over \mu_\ell} 
\big(e_i -\mu_\ell\big)~
=~\sqrt{\mu_{\ell+1}\mu_m\over \mu_\ell \mu_{m+1}}
\times \sqrt{\mu_{m+1} \over \mu_m}(e_i -\mu_m) +\sqrt{\mu_{\ell+1}
\over \mu_{\ell}} (\mu_m-\mu_\ell) }
where we have reconstructed a white box addition at point
$i$ and height $m$ in the first term, we simply get
\eqn\actup{ 
\sqrt{\mu_{\ell+1} \over \mu_\ell} 
\big(e_i -\mu_\ell\big) (b)_2 ~=~ 
\sqrt{\mu_{\ell+1}\mu_m\over \mu_\ell \mu_{m+1}} (b+\diamond_{i,m})_2+
\sqrt{\mu_{\ell+1}\over \mu_\ell}(\mu_m-\mu_\ell) (b)_2 }
\par                                                     
These three situations are summarized in the following recursion
relation, according to the configuration of $b$ at $i$, respectively
denoted $\delta_{b,max(i,m)}$ (case (1)) 
and $\delta_{b,min(i,m)}$ (case (3))
\eqn\recuprot{\eqalign{
\sqrt{\mu_{\ell+1}\over \mu_\ell}(e_i-\mu_\ell)(b)_2~&=~
-\sqrt{\mu_{\ell+1}\mu_\ell}\  (b)_2 \cr
&+\delta_{b,max(i,m)}
\bigg[
\sqrt{\mu_{\ell+1} \over \mu_\ell}\mu_{m+1}^{-1} (b)_2+
\sqrt{\mu_{\ell+1}\mu_{m} \over \mu_\ell \mu_{m+1}}
(b-\diamond_{i,m})_2\bigg] \cr
&+\delta_{b,min(i,m)}\bigg[ 
\sqrt{\mu_{\ell+1}\mu_m\over \mu_\ell \mu_{m+1}} (b+\diamond_{i,m})_2+
\sqrt{\mu_{\ell+1}\over \mu_\ell}\mu_m (b)_2\bigg] \cr}}
In all cases (1-3), this enables us to 
reexpress 
\eqn\reexd{ (a+\diamond_{i,\ell})_2^t (b)_2~=~(a)_2^t 
\sqrt{\mu_{\ell+1}\over \mu_\ell}(e_i-\mu_\ell) (b)_2~=~\sum_{b'\in
W_{2n}} \lambda_{b'} (a)_2^t (b')_2}
as a linear combination of terms of the form $(a)_2^t (b')_2$, where
$b'=b$, $b+\diamond$ or $b-\diamond$, hence with $|b'|\geq |b|-1$.
But, by hypothesis \hypolen, we have $|b|\geq|a|+1$, hence 
$|b'|\geq|a|$.
We can therefore apply the recursion hypothesis $\cal P$ \prorec\
to each of the products $(a)_2^t (b')_2$ in \reexd, 
which must then vanish,
and we finally get a zero answer for $(a+\diamond)_2^t (b)_2$.
This establishes the property $\cal P$ \prorec\ for $a+\diamond$,
under the assumption that it is satisfied for $a$.
\par
To complete the recursion, we have to establish the property
$\cal P$ \prorec\ for the initial point $a=a_n$. It will then 
hold for any $a \in W_{2n}$.
Let us prove that
\eqn\initrec{ (a_n)_2^t (b)_2~=~ 0 \qquad{\rm for} \ {\rm all}\ b\in
W_{2n}\ {\rm such}\ {\rm that} \ b \neq a_n}
(The condition that $|b|\geq |a_n|=0$ does not give any 
restriction on $b$.).
We have to  
evaluate the product $e_1 e_3 ... e_{2n-1} (b)_2$, namely the
addition of a row of $n$ grey boxes to $b$. 
As those cover the whole width of $b$, there is at least one such grey box
which acts on a white slope of $b$ (otherwise, $b$ should
have no slope, hence would be equal to $a_n$).
But in eqs \greybw-\nediagr\
above, we have proved that the addition of a grey box on a white
slope of $(b)_2$ yields a zero answer. 
This completes the proof of \initrec, and the lemma 1 follows
by recursion.
\par
To prove the proposition 1, we note that by symmetry of the bilinear form
\bilin\ 
\eqn\symbil{\eqalign{ 
\big((a)_2,(b)_2\big)~&=~{\rm Tr}\big( (a)_2 (b)_2^t\big)
~=~{\rm Tr}\big( (b)_2^t (a)_2\big)\cr
&=~\big( (b)_2,(a)_2\big)~=~{\rm Tr}\big( (a)_2^t (b)_2 \big)\cr}}
If $a\neq b$ we immediately get $\big( (a)_2, (b)_2\big)=0$
by applying the lemma 1 to $(a,b)$ if $|a|<|b|$ or $(b,a)$
if $|a|>|b|$, and either of the two if $|a|=|b|$.
This gives the orthogonality of the distinct basis 2 elements
w.r.t. the bilinear form \bilin.
\par
The norm of $(a)_2$, $a\in W_{2n}$, 
is easily computed by recursion. 
We have already seen in \norinit\ that the
fundamental basis 2 element has norm $\big((a_n)_2,(a_n)_2\big)=1$.
Suppose that $\big( (a)_2,(a)_2\big)=1$ for some $a \in W_{2n}$.
Let us compute the norm of the element $(a+\diamond)_2$,
for some white box addition on $a$.
We have
\eqn\calnor{\eqalign{
(a+\diamond_{i,\ell})_2^t (a+\diamond_{i,\ell})_2~&=~
(a)_2^t {\mu_{\ell+1}\over \mu_\ell} (e_i -\mu_\ell)^2 (a)_2 \cr
&=~(a)_2^t \bigg( 1+ \sqrt{\mu_{\ell+1}\over \mu_\ell}
(\mu_{\ell+1}^{-1}-\mu_\ell)\sqrt{\mu_{\ell+1}\over \mu_\ell}
(e_i-\mu_\ell)  \bigg) (a)_2 \cr
&=~
(a)_2^t (a)_2+
\sqrt{\mu_{\ell+1}\over \mu_\ell}
(\mu_{\ell+1}^{-1}-\mu_\ell) (a)_2^t (a+\diamond_{i,\ell})_2 
\cr
&=~(a)_2^t(a)_2\cr}}
where we have used $e_i^2=\mu_1^{-1}e_i$ and the relation \murela\
in the first line, and $(a)_2^t (a+\diamond)_2=0$ 
in the second, by direct application of
the lemma 2.
Eq.\calnor\ implies that 
$\big( (a+\diamond)_2,(a+\diamond)_2\big)=\big( (a)_2,(a)_2\big)=1$,
by the recursion hypothesis. Together with the initial point
\norinit, this proves that $\big( (a)_2,(a)_2\big)=1$,
for all $a\in W_{2n}$. The proposition 1 follows.
\par
\subsec{The meander determinant}
The meander determinant \detgra\ follows from
the Gram determinant for the basis 1. Let us now compute the latter.
The basis 2 being orthonormal, its Gram matrix is the $c_n\times c_n$
identity matrix $I$. The change of basis from basis 1 to 2 (with 
the upper triangular matrix $P$ \chbas)
therefore reads
\eqn\chbmat{ P\Gamma_{2n}(q)P^t~=~  I }
Hence we have $\det \Gamma_{2n}(q)=(\det P)^{-2}$.
As $P$ is an upper triangular matrix, only the diagonal elements
$P_{a,a}$ enter the determinant formula.
From the definition of the basis 2 elements by white box additions
\boxrule\
on the fundamental $(a_n)_2$, we immediately get that
the matrix elements $P_{a,a}$ satisfy the recursion
\eqn\recPmat{ P_{a+\diamond_{i,\ell},a+\diamond_{i,\ell}}~=~
\sqrt{\mu_{\ell+1}\over \mu_{\ell}} P_{a,a} }
With the initial condition $P_{a_n,a_n}=\mu_1^n$ for the fundamental
walk diagram, this determines the $P_{a,a}$ completely.
We have 
\eqn\pmatelem{ P_{a,a}^2~=~\mu_1^{2n}\prod_{{\rm boxes}\ {\rm of}\ a}
{\mu_{\ell+1} \over \mu_\ell} }
\fig{The decomposition of a walk $a\in W_{12}$ into strips of white boxes.
There are $n=6$ such strips, with respective lengths $2$, 
$4$, $3$, $2$, $1$
and $1$ (note that an empty strip has by definition 
length $1$).}{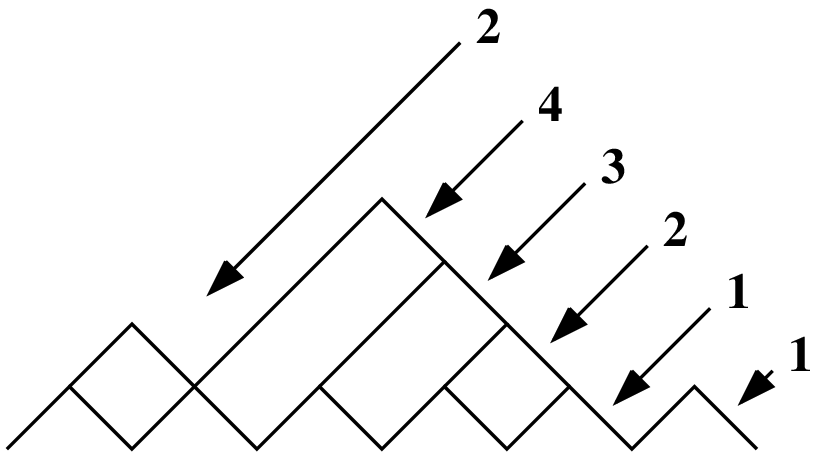}{6.cm}
\figlabel\strips
The boxes of any $a\in W_{2n}$ can be arranged into $n$ {\it strips},
as illustrated in Fig.\strips, namely $n$ diagonal lines
of boxes of increasing
consecutive heights and positions. Each such line has
an upper end, the top of the rightmost box in the line. Let
us call the height of this end the length of the corresponding strip.
For instance, the fundamental diagram is formed of $n$ 
strips of length $1$.
With this definition, we simply get
\eqn\finP{ P_{a,a}^2~=~ \mu_1^n \prod_{{\rm strips}\ {\rm of}\ a}
\mu_\ell }
where, in the product over the $n$ strips of $a$, 
$\ell$ stands for the corresponding strip length.
The Gram determinant of the basis 1 reads then
\eqn\gradeter{ \det \Gamma_{2n}(q)~=~\prod_{a \in W_{2n}} P_{a,a}^{-2}~
=~\mu_1^{-nc_n} \prod_{{\rm strips}\ {\rm of}\atop
{\rm all}\ a\in W_{2n}} \mu_\ell^{-1} }
We also get a formula for the meander determinant, using \detgra\
\eqn\medetmu{ \det {\cal G}_{2n}(q) ~=~
\prod_{{\rm strips}\ {\rm of }\atop 
{\rm all}\ {\rm walks}\ \in W_{2n}} 
\mu_{\ell}^{-1} } 
\par
This can be rewritten as
\eqn\rewdetfor{ \det {\cal G}_{2n}(q)~=~ \prod_{m=1}^n 
\big[\mu_m\big]^{-s_{2n,m}} }
where $s_{2n,m}$ denotes the total number of strips of length
$m$ in all the walk diagrams of order $2n$, $W_{2n}$.  
The formula of theorem 1 will follow from the explicit computation of
the numbers $s_{2n,m}$. We have the
\par
\deb
\noindent{\bf PROPOSITION 2:}
\par
\eqn\snjcal{ s_{2n,m}~=~c_{2n,2m}~=~{2n \choose n-m}-{2n \choose n-m-1} }
\fin
\par
This will be proved by establishing a bijection
between the walks $a\in W_{2n}$ {\it with a marked end of strip} at
height $m$, and the walks of $2n$ steps on a half-line starting
at the origin $h(0)=0$ and ending at height $h(2n)=2m$, namely
the elements $b\in W_{2n}^{(2m)}$. The cardinal of the latter set 
being equal to $|W_{2n}^{(2m)}|=c_{2n,2m}$ (see eq.\bijwa), the 
proposition 2 will follow.
\par
\fig{The map from any
$a\in W_{2n}^{(2m)}$ to a walk  $b \in W_{2n}$
with a marked end of strip at height $m$. 
$i$ is the rightmost intersection of the line
$h=m$ with $a$ at an ascending slope. The walk $a$ is cut into two 
parts: the left $L\in W_{i}^{(m)}$, 
and the right $R$, such that its reflection
$\bar R \in W_{2n-i}^{(m)}$. We have $b=L\bar R\in W_{2n}$, with the marked
point at $i$. If $h(i+1)=m-1$ (not the case in the present figure),
this is the desired walk of $W_{2n}$ with a marked end of strip
at height $m$. If $h(i+1)=m+1$ (the case of the present figure), 
we migrate 
$i \to i'=$min$\{j>i|h(j)=m=h(j+1)+1\}$, and mark $i'$.
The migration is indicated by an arrow. The corresponding strip 
of length $m$ 
has also been represented.}{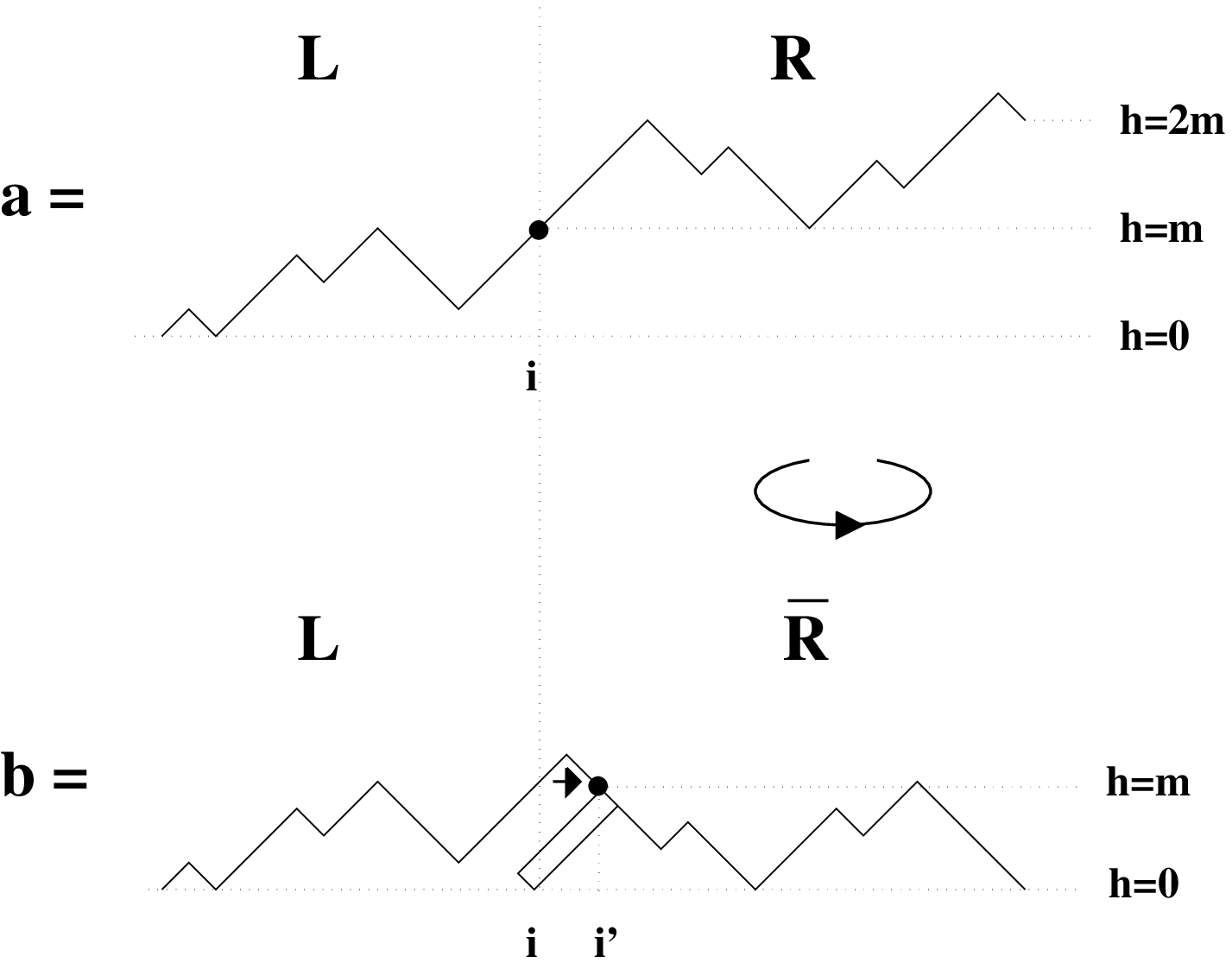}{10.cm}
\figlabel\refletran
Let us consider any walk $a\in W_{2n}^{(2m)}$.
As shown in Fig.\refletran,
the line $h=m$ intersects the walk $a$ at least once along an 
ascending slope (at some point $j$ where $h(j)=m$ and $h(j-1)=m-1$
on $a$). 
Let $i$ denote the position of the 
{\it rightmost}\foot{The fact that we take the rightmost intersection 
here is responsible for the bijectivity of the mapping.}
such intersection, namely $i=$max$\{j|h(j)=m=h(j-1)+1\}$. 
Cutting
the walk $a$ at the point $(i,h(i)=m)$ separates the walk into
a left part $L\in W_{i}^{(m)}$ and a right part $R$,
which may be viewed as an element of $W_{2n-i}^{(m)}$ (see Fig.\refletran).
Indeed, from the definition of $i$, the walk $R$ stays above the line
$h=m$ until its end: subtracting $m$ from all its heights,
and counting its steps from 0 to $2n-i$ (instead of from $i$ to $2n$)
expresses $R$ as an element of $W_{2n}^{(m)}$. 
Reflecting $R \to \bar R$, i.e. describing it in the opposite direction
($\bar R$ is a walk on the half-line starting at height $m$ and
ending at height $0$ after $2n-i$ steps), and composing $L$ and
$\bar R$, i.e. attaching the origin of $\bar R$ to the end
of $L$, we form a walk $b=L\bar R \in W_{2n}$ (see Fig.\refletran). 
In this walk, we have $h(i)=m$. If $h(i+1)=m-1$, $i$ is an end of strip
of height $m$, which we mark. If $h(i+1)=m+1$, $i$ cannot be an 
end of strip. Nevertheless, we just have to consider the 
smallest point $i'>i$ such that $h(i')=m=h(i'+1)+1$, which always
exists, as the walk $a$ goes back to height 0 at position $2n$. 
This point $i'$ is an end of strip at height $m$, which we mark.
\par
Conversely, let us
start from some $a\in W_{2n}$ with a marked end of strip at
position $i$ and height $m$.  By definition, this end of strip satisfies 
$h(i)=m$ and $h(i+1)=m-1$. If $i$ is a maximum of $a$, 
namely $h(i-1)=m-1$,
it  separates the walk $a$ into
a left part $L$ and a right part $R$. The left part is a walk
on the half-line, ending at height $m$ after $i$ steps, hence
$L\in W_{i}^{(m)}$. The right part $R$ is a walk on the half-line
starting at height $m$ and ending at the origin, after $2n-i$
steps. 
The reflected walk $\bar R$ is obtained by describing $R$ in the
opposite direction, namely starting from the origin, and ending at
height $m$, after $2n-i$ steps. Hence we can 
write that $\bar R\in W_{2n-i}^{(m)}$.
Now if we compose the walks $L$ and $\bar R$ (attach the origin of $\bar R$
to the end of $L$), the resulting walk $b=L\bar R\in W_{2n}^{(2m)}$,
and due to the fact that $i$ was a maximum of $a$, we have 
$h(2n-1)=2m-1$ and $h(2n)=2m$ in $b$.
If $i$ is not a maximum of $a$, we first migrate the marked point
from $i$
to the largest value $i'<i$, such that $h(i')=m=h(i'-1)+1$
(the closest ascending slope at height $m$ to the  left of $i$).
Then we apply the previous cutting, reflecting and pasting procedure
at the point $i'$. This produces a walk $b=L\bar R \in W_{2n}^{(2m)}$,
with the particular property that $h(2n-1)=2m+1$ and $h(2n)=2m$ 
on $b$. 
\par 
We have in fact established a more refined mapping
between (i) the $a\in W_{2n}$ with a marked maximum, 
of height $m$ (namely at a point $i$
such that $h(i)=m=h(i+1)+1=h(i-1)+1$) and the 
$b\in W_{2n-1}^{(2m-1)}$ (ii) the $a \in W_{2n}$ with a marked 
descending slope at height $m$ ($i$ such that $h(i)=m=h(i-1)-1=h(i+1)+1$)
and the $b\in W_{2n-1}^{(2m+1)}$.
This forms a bijection between the walks $a\in W_{2n}$ with a marked
end of strip (either a maximum or a descending slope) and the walks
$b\in W_{2n}^{(2m)}$ (with either $h(2n-1)=2m-1$ or $h(2n-1)=2m+1$).
Hence we conclude that
\eqn\concluop{ s_{2n,m}~=~|W_{2n}^{(2m)}|~=~c_{2n,2m} }
which proves the proposition 2.
\par
To translate the result \snjcal\ of proposition 2 into the
formula of theorem 1, using \rewdetfor, we simply have to 
reexpress the meander determinant in terms of the Chebishev
polynomials $U_m(q)$, using $\mu_m=U_{m-1}/U_m$.
Eq \rewdetfor\ becomes
\eqn\rexpdet{ \det {\cal G}_{2n}(q)~=~\prod_{m=1}^n \bigg(
{U_m\over U_{m-1}}\bigg)^{s_{2n,m}}~=~\prod_{m=1}^n 
\big[U_m\big]^{s_{2n,m}-s_{2n,m+1}}}
by noting that $s_{2n,n+1}={2n \choose -1}-{2n \choose -2}=0$.
This takes exactly the form of \thone, with
\eqn\evalfinone{ a_{2n,2m}~=~s_{2n,m}-s_{2n,m+1}~=~c_{2n,2m}-c_{2n,2m+2}}
which completes the proof of the theorem 1.
\par
\newsec{The semi-meander determinant: proof of theorem 2} 
The strategy of the proof of theorem 2 is exactly the same
as for theorem 1. 
It is based on the representation of open arch configurations
by a particular set of reduced elements of the Temperley-Lieb
algebra, forming the basis (still called basis 1, but not to
be confused with that of previous section) 
of a vector subspace thereof.
The semi-meander determinant is then expressed in terms
of the Gram determinant of this basis 1. The next step
is the explicit
Gram-Schmidt orthogonalization of this basis, defining
another basis, called basis 2. The semi-meander determinant
is then computed by using the change of basis 1 $\to$ 2.
\par
\subsec{Temperley-Lieb algebra and open arch configurations} 
The open arch configurations of $A_n^{(h)}$, with
order $n$ and with $h$ open arches, can be represented
by some particular reduced elements of the Temperley-Lieb
algebra $TL_n(q)$. 
\fig{The interpretation of an open 
arch configuration of order $n=15$
and with $h=3$ open arches (right diagram) as a reduced element
of $TL_{15}(q)$ (left diagram). 
Note that exactly $h=3$ strings go across the domino,
namely link three lower to (the three rightmost)
upper ends. The linking of the upper ends
of the domino is made through $(n-h)/2=6$ strings connecting consecutive
ends by pairs.}{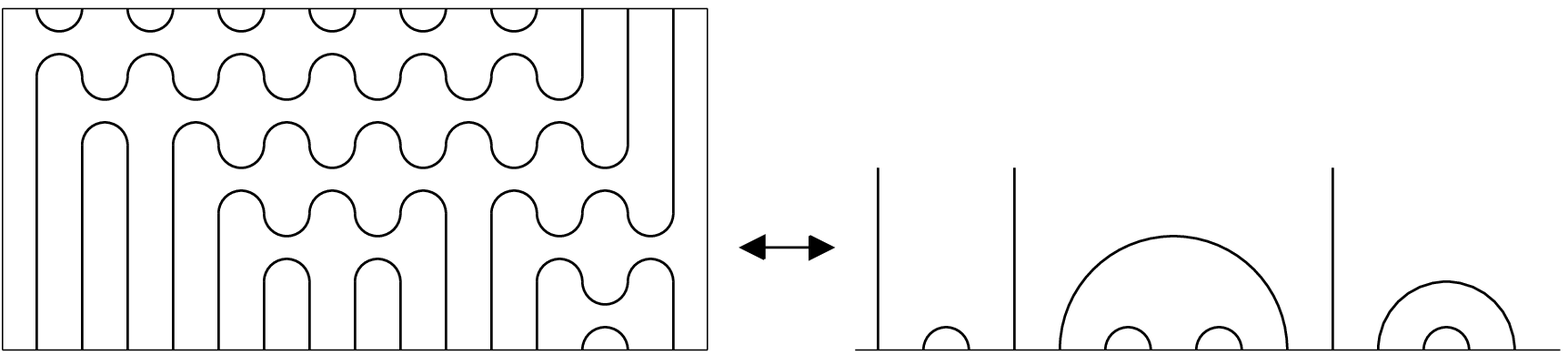}{10.cm}
\figlabel\opred
In Fig.\opred, we have represented in the string-domino pictorial 
representation the domino corresponding to a reduced 
element of the Temperley-Lieb algebra, immediately interpretable
as an open arch configuration. 
Starting from $a\in A_n^{(h)}$, let us construct an element,
still denoted\foot{Here we adopt the same notation for elements 
of $TL_n(q)$ corresponding to open arch configurations as that used
before for closed arch configurations. 
These will correspond to another basis $\{ (a)_1\}$ for $a\in A_n^{(h)}$,
which we will refer to again as the basis 1. This should
not be confusing, as we are only dealing with the open arch case from now on.}
by $(a)_1$ of $TL_n(q)$:
representing the corresponding domino as acting
from bottom to top, the connection of
its $n$ lower ends of strings is realized through the closed arches
of $a$, whereas the $h$ open arches just go across the domino,
and connect $h$ of the lower ends to the $h$
rightmost upper ends of strings. The remaining $n-h$ ends are 
then connected by
consecutive pairs like in the meander case.
This construction establishes a bijection between $A_n^{(h)}$ 
and the reduced elements of $TL_n(q)$ with exactly $h$ strings
connecting lower ends to the $h$ rightmost upper ends, and $(n-h)/2$
strings connecting the remaining $n-h$ upper ends by consecutive pairs.
Let us denote by ${\cal I}_n^{(h)}(q)$ the vector space spanned by
these reduced elements. 
From now on, we will refer to the basis $\{ (a)_1|a\in A_n^{(h)}\}$ 
as the basis 1.
\par
Like in the meander case, the basis 1 is best expressed in the 
equivalent language
of walk diagrams $a\in W_n^{(h)}$.
Let $a_n^{(h)}$ be the fundamental element of $W_n^{(h)}$, with
$h(0)=h(2)=...=h(n-h)=0$, 
$h(1)=h(3)=...=h(n-h-1)=1$ 
and $h(n-h+j)=j$ for $j=1,2,...,h$.
Any $a\in W_n^{(h)}$ may be viewed as the result of box additions
on the fundamental $a_n^{(h)}$. The construction of $(a)_1$,
$a \in W_n^{(h)}$ is performed recursively.
We first set
\eqn\firelsd{ (a_n^{(h)})_1~=~e_1 e_3 ... e_{n-h-1} }
and then for a box addition at position $i$, we set
\eqn\boxsd{ (a+\diamond_i)_1~=~ e_i (a)_1 }
As an example, the basis 1 elements for ${\cal I}_4^{(2)}$
read
\eqn\foutwo{\eqalign{
\bigg(\figbox{1.2cm}{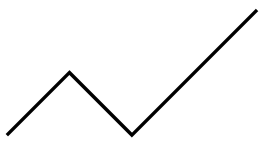}\bigg)_1~&=~e_1\qquad 
\bigg(\figbox{1.2cm}{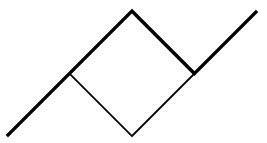}\bigg)_1~=~e_2 e_1\cr
\bigg(\figbox{1.2cm}{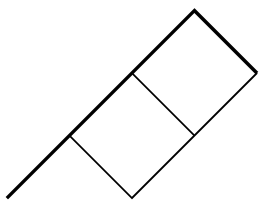}\bigg)_1~&=~e_3 e_2 e_1\cr}}
where we have represented the boxes added on the walk diagrams.
\par
\fig{The string-domino picture corresponding to 
the box decomposition of an open walk diagram $a\in W_{11}^{(3)}$.
Note that exactly $3$ strings join upper and lower ends.
The domino is rather read from top to bottom, as opposed to the case
of Fig.\opred, where it is read from bottom to top.}{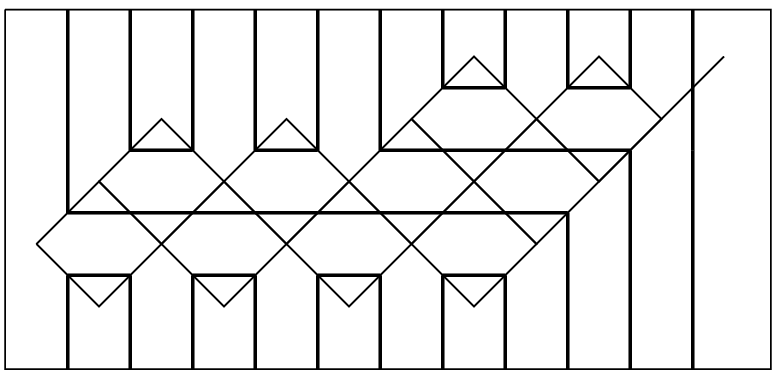}{6.cm}
\figlabel\dostri
To make direct contact with the string-domino pictorial representation,
we may attach to the box decomposition of any walk diagram $a\in
W_n^{(h)}$ a domino using the same rule as in Sect.3.1, namely 
represent {\it all } the boxes corresponding to left multiplications by
$e_i$ (including those of the fundamental element $a_n^{(h)}$),
and decorate them by a horizontal double line 
(string), as in \symbox. The picture is then completed by drawing
vertical strtings joining the string ends on the upper and lower borders
of the domino. This is illustrated in Fig.\dostri, where the strings are
represented
in thick black lines. 
\par
The main and new difficulty here, in comparison with the former meander
case, is that these reduced
elements of $TL_n(q)$ {\it do not} form an ideal\foot{This will be
responsible for the
absence of a generalization of the lemma 1 of Sect.3.3 for the
present case.}. 
For instance, we have listed in \foutwo\ the basis 1 elements
for ${\cal I}_4^{(2)}(q)$. If we multiply the first (fundamental)
element by the third one, we find
$(e_1)(e_3 e_2 e_1)=e_1 e_3$  
which does not belong to the space ${\cal I}_4^{(2)}(q)$
(there is no string connecting lower and upper ends in $e_1 e_3$,
whereas there must be 2 such strings 
in any element of ${\cal I}_4^{(2)}(q)$),
which is therefore not an ideal.
\par
Nevertheless, we can still form the Gram matrix $\Gamma_n^{(h)}(q)$
for the basis 1, by using
the restriction to ${\cal I}_n^{(h)}(q)$ of the bilinear form
\bilin. This reads
\eqn\grasd{ \big[ \Gamma_n^{(h)}(q)\big]_{a,b}~=~
\big( (a)_1,(b)_1\big) \qquad {\rm for} \ a,b \in A_n^{(h)} }
\fig{Computation of $\big((a)_1,(b)_1\big)$.
We put the reflected domino $(b)_1^t$ on top of the domino
$(a)_1$ (here, $a,b \in W_{11}^{(3)}$). The upper ends
are then identified one by one to the lower ends of strings.
Counting the loops formed yields: $(n-h)/2=4$ central loops
formed at the connection between the two dominos, plus
$\kappa(a|b)=3$ loops coming from the superposition of the
open arch configurations $a$ and $b^t$ (reflected w.r.t. the river).
This gives finally $\big( (a)_1,(b)_1\big)=q^{7}$.}{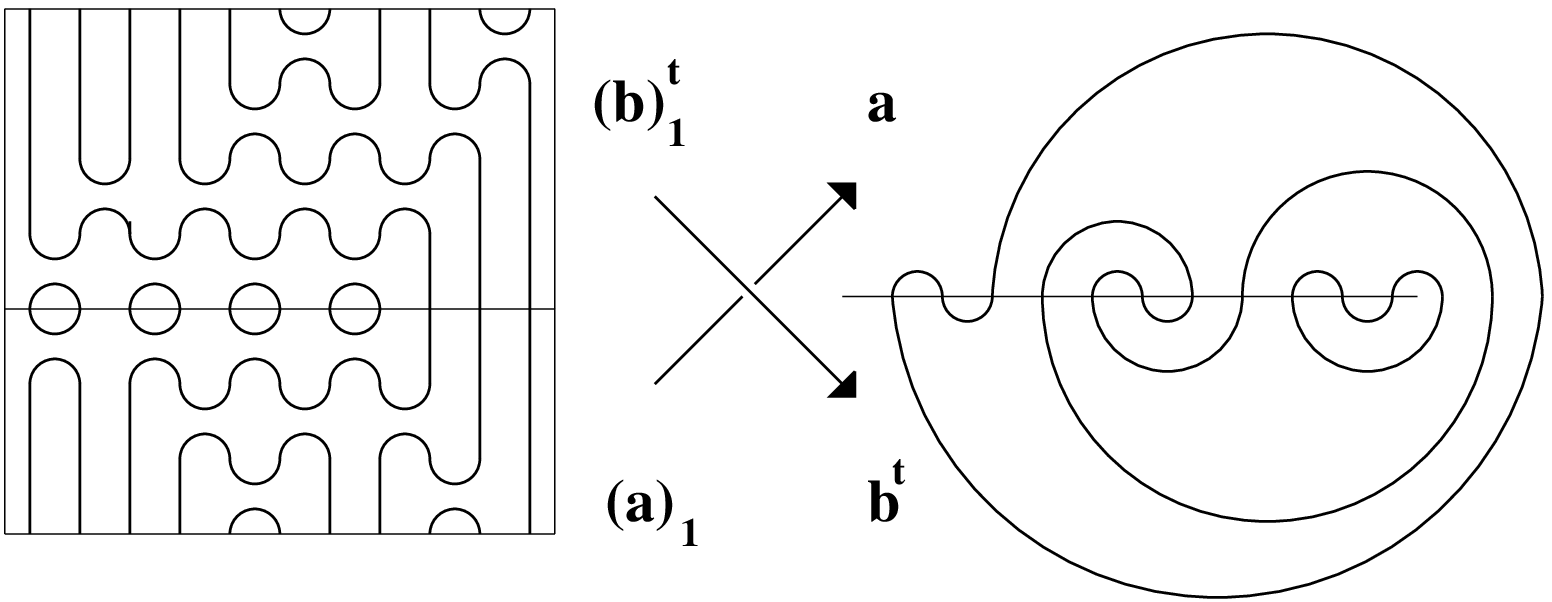}{10.cm}
\figlabel\semdom
As illustrated in Fig.\semdom,
to compute $\big( (a)_1,(b)_1\big)$, we glue the dominos
$(a)_1$ and the reflected $(b)_1^t$, identify the upper and lower string
ends, and count the number of resulting connected components.
The connection of the two dominos creates $(n-h)/2$ loops, from
the strings connecting the upper ends by consecutive 
pairs on $(a)_1$ and $(b)_1$.
The remaining part simply creates $\kappa(a|b)$ loops, from the
superposition
of the open arch configurations $a$ and $b^t$ (reflected w.r.t. the river),
and the connection of their $h$ open arches (see Fig.\semdom).
Hence the Gram matrix for the basis 1 of ${\cal I}_n^{(h)}(q)$
is simply related to the semi-meander matrix \mesemat, through
\eqn\linkgasd{ [\Gamma_n^{(h)}(q)\big]_{a,b}~=~ 
q^{{n-h\over 2}+\kappa(a|b)}~=~
q^{n-h\over 2} \big[{\cal G}_n^{(h)}(q)\big]_{a,b} }
The semi-meander determinant is therefore related to the
Gram determinant of the basis 1 through
\eqn\relsdetg{ \det {\cal G}_n^{(h)}(q)~=~
\mu_1^{{n-h \over 2} c_{n,h}} \det \Gamma_n^{(h)}(q)}
\par
\subsec{Orthogonalization of the basis 1} 
In this section, we introduce a basis 2 of ${\cal I}_n^{(h)}(q)$,
still indexed by $a\in W_n^{(h)}$, which will be orthonormal 
with respect to the bilinear form \bilin.
\par
Like in the meander case, the basis 2 will be defined recursively
through box additions.  We start from the basic element
\eqn\basd{(a_n^{(h)})_2~=~\mu_1^{n/2} (a_n^{(h)})_1} 
where the normalization ensures 
that $\big( (a_n^{(h)})_2,(a_n^{(h)})_2\big)=
q^{-n} q^{n-h\over 2} q^{n+h\over 2}=1$, where we have counted the  
contributions of the $(n-h)/2$ loops formed by the strings pairing
upper ends by consecutive pairs on $(a)_1$, and that of the 
$\kappa(a|a)=(n+h)/2$ loops created by the superposition of $a$ with 
its own reflection $a^t$.
\par
To proceed, we need to define the concept of {\it floor}
of a walk diagram $a\in W_n^{(h)}$. Let us denote by $h(i)$, 
$i=0,1,2,...,n$ the heights of $a$, with $h(0)=0$ and $h(n)=h$.
The floor of $a$ is yet another diagram $f(a)\in W_n^{(h)}$,
such that $f(a)\subset a$, and with heights $h'(i)$, $i=0,1,2,...,n$,
defined  as follows.
Let us denote by $J$ the set 
of integers
\eqn\jdef{ J~=~ \{ j \in \{0,1,...,n\} \ {\rm such}\ {\rm that}\ 
h(k)\geq h(j)\ , \forall \ k\geq j\} }
\fig{A diagram $a\in W_{20}^{(6)}$ (thick black line) 
and the construction
of its floor $f(a)\subset a$. The segments $J_0$, $J_1$, ..., $J_4$
of positions forming $J$
are indicated by dotted lines. 
The floor $f(a)$ is represented
filled with grey boxes.
The boxes inbetween $f(a)$ and $a$ are
represented
in white. The floor-ends have positions
$0$, $4$, $6$, $12$, $14$, $16$, $17$, $19$, 
$20$.}{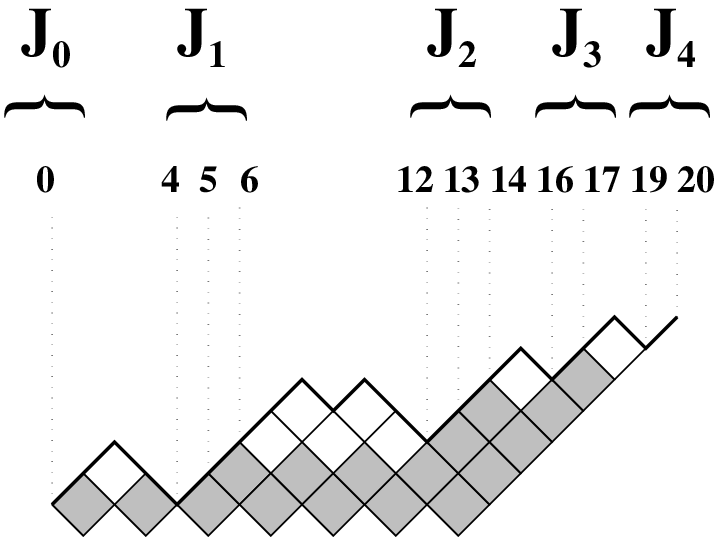}{6.cm}
\figlabel\floflo
As illustrated in Fig.\floflo,
this set $J$ is clearly the union of ordered segments of positions,
of the form $j=J_0\cup J_1 \cup ... \cup J_k$, with
$J_i=\{ j_i, j_i+1, j_i+2,...,j_i+n_i\}$, for some integers $n_i$
and $j_i$, $i=0,...,k$. These segments correspond to the ascending
slopes of $a$ such that no point on their right 
has a lower height.
With these notations, the floor $f(a)$ 
of $a$ is defined to have the heights
$h'(j)$, $j=0,1,...,n$, according to the following rules
\eqn\flohei{
\eqalign{  h'(j)~&=~h(j) \qquad \forall \ j \in J \cr
h'(j_i+n_i+2r)~&=~h'(j_i+n_i) \qquad \forall\ r\geq 0
\ {\rm with}\ 2r \leq j_{i+1}-j_i-n_i\cr
h'(j_i+n_i+2r-1)~&=~h'(j_i+n_i)-1 \qquad \forall 
\ r \geq 1\ {\rm with}\ 2r-1 \leq (j_{i+1}-j_i-n_i)\cr}}
This is valid for all $i=1,2,...,k$. For $i=0$, we have to be more
careful, as the leftmost floor
piece has a different status. 
If $J_0\neq \{0\}$ (this leftmost floor piece is empty), 
then \flohei\ is valid for $i=0$ as well.
If $J_0=\{0\}$ (this leftmost floor piece is not empty:
this is the case in Fig.\floflo),
we have to add the values
\eqn\adeqg{\eqalign{
h'(0)~&=~h'(2)~=~\cdots ~=~h'(j_1)~=~0\cr
h'(1)~&=~h'(3)~=~\cdots ~=~h'(j_1-1)~=~1\cr}}   
The floor diagram is represented filled with grey 
boxes in Fig.\floflo.
The floor diagram $f(a)$ is in fact a succession of horizontal
broken lines, with heights alternating $h(j_i+n_i)=\ell+1$, 
$h(j_i+n_i+1)=\ell$, $h(j_i+n_i+2)=\ell+1$,..., $h(j_{i+1})=\ell+1$,
on the intermediate positions inbetween the segments $J_i$ and
$J_{i+1}$. These are separated by ascending
slopes (along the segments $J_i$). For each such intermediate floor
$F_i$,
we define the {\it floor height} to be the number
$\ell=h(j_i+n_i-1)=h(j_i+n_i+1)=...=h(j_{i+1})-1$, for $i\geq 1$.
The leftmost floor $F_0$, of height 0 if $J_0=\{0\}$, 
is a little different as we have
$\ell=0=h(j_0=0)=h(2)=...=h(j_1)$ from \adeqg.
We will also refer to these intermediate floors as simply the floors
of $a$, for which this decomposition is implied. The endpoints
with positions
$j_i+n_i$ and $j_{i+1}$ (and equal height $h(j_i+n_i)=h(j_{i+1})$
except maybe for the rightmost floor-end)
of each of these floors will be called floor-ends in the following.
\par
To define the basis 2 of ${\cal I}_n^{(h)}(q)$,  
we will need a pictorial representation of the walk diagrams
$a\in W_n^{(h)}$ in which the floor $f(a)$ is also represented.
As in Sect.3, we adopt the representation \notboru\ by grey
and white boxes  of the left mutliplications
of a reduced element of ${\cal I}_n^{(h)}(q)$ by respectively
$e_i$ at position $i$ or $\sqrt{\mu_{m+1}/\mu_m} (e_i-\mu_m)$
on a minimum of height $m$ and position $i$.
The basis 2 elements then correspond to 
\item{(i)} grey box additions for all the boxes forming the floor
$f(a)$, including the basic boxes forming $a_n^{(h)}$ (see below) 
\item{(ii)} white box additions for all the superstructures of
$a$ above its floor $f(a)$. There is however
a final subtlety with the height of these white boxes, which
is counted along strips, w.r.t. the grey floor. 
\par
\noindent{}
In the case \basd\ of the fundamental diagram, the representation
is simply
\eqn\repbas{\eqalign{\big( a_n^{(h)}\big)_2~&=~ 
\mu_1^{n/2} \figbox{4.cm}{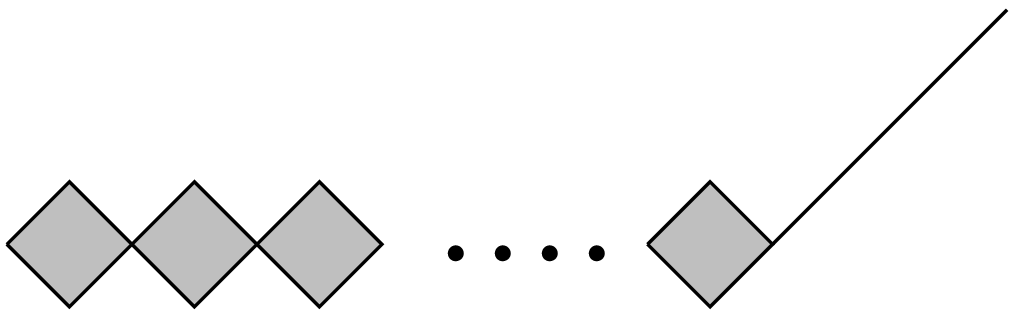} \cr
&=~\mu_1^{n/2} e_1 e_3 ... e_{n-h-1} \cr}}
as the floor of this element
is simply $f(a_n^{(h)})=a_n^{(h)}$, and we have represented the basic
grey boxes under the floor.
The other elements of the basis 2 are obtained by white box additions on
$\big(a_n^{(h)}\big)_2$.
The novelty, when compared to the case of Sect.3, is that some box
additions may create a new floor, namely change previously
added white boxes into grey ones.  
\par
In general, the best way to construct the basis 2 elements,
is to first list all the walk diagrams $a\in W_n^{(h)}$,
represent them together with their floor $f(a)\subset a$, and
then write the corresponding products of grey and white boxes.
This is illustrated now in the case of $W_6^{(2)}$. 
\eqn\sixtwo{\eqalign{
\bigg(\figbox{1.5cm}{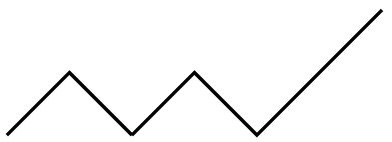}\bigg)_2~&=~\mu_1^3\figbox{1.5cm}{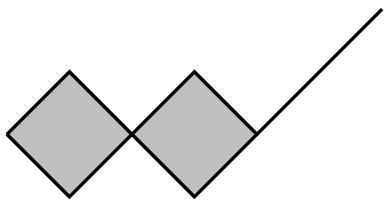}
~=~ \mu_1^3 e_1 e_3 \cr
\bigg(\figbox{1.5cm}{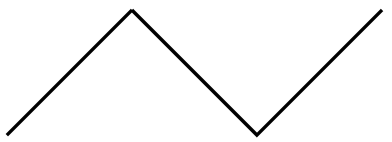}\bigg)_2~&=~\mu_1^3\figbox{1.5cm}{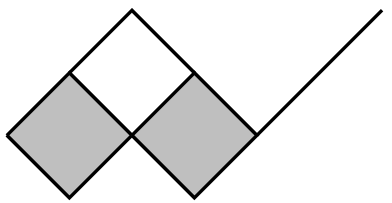}
~=~ \mu_1^{5/2} \mu_2^{1/2}(e_2-\mu_1) e_1 e_3 \cr
\bigg(\figbox{1.5cm}{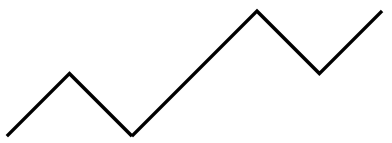}\bigg)_2~&=~\mu_1^3\figbox{1.5cm}{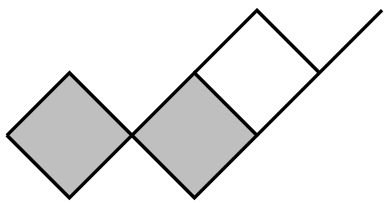}
~=~ \mu_1^{5/2} \mu_2^{1/2}(e_4-\mu_1) e_1 e_3 \cr
\bigg(\figbox{1.5cm}{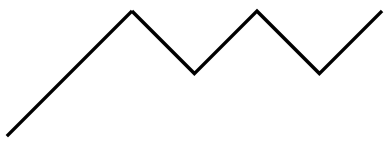}\bigg)_2~&=~\mu_1^3\figbox{1.5cm}{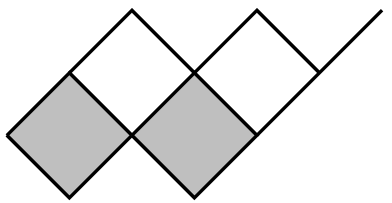}
~=~ \mu_1^2 \mu_2(e_2-\mu_1)(e_4-\mu_1) e_1 e_3 \cr
\bigg(\figbox{1.5cm}{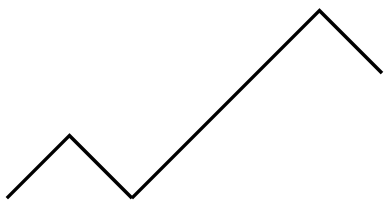}\bigg)_2~&=~\mu_1^3\figbox{1.5cm}{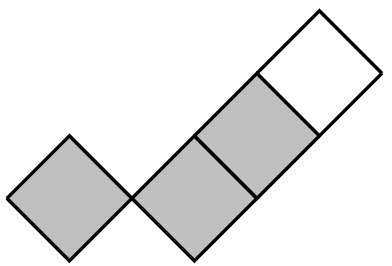}
~=~ \mu_1^{5/2} \mu_2^{1/2}(e_5-\mu_1) e_4 e_1 e_3 \cr
\bigg(\figbox{1.5cm}{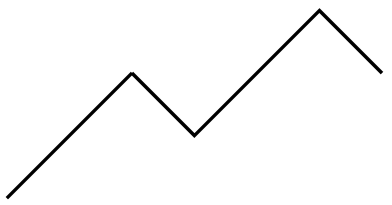}\bigg)_2~&=~\mu_1^3\figbox{1.5cm}{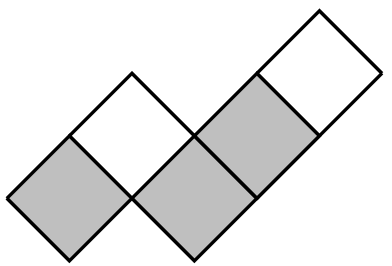}
~=~ \mu_1^2 \mu_2(e_2-\mu_1)(e_5-\mu_1) e_4 e_1 e_3 \cr
\bigg(\figbox{1.5cm}{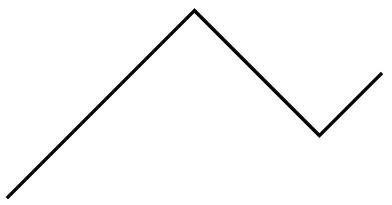}\bigg)_2~&=~\mu_1^3\figbox{1.5cm}{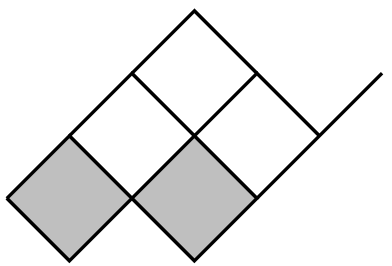} \cr
&=~ \mu_1^2\mu_2^{1/2}\mu_3^{1/2}
(e_3-\mu_2)(e_2-\mu_1)(e_4-\mu_1) e_1 e_3 \cr
\bigg(\figbox{1.5cm}{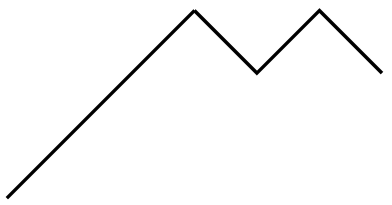}\bigg)_2~&=~\mu_1^3\figbox{1.5cm}{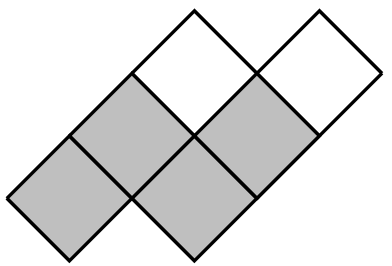}
~=~ \mu_1^2 \mu_2(e_3-\mu_1)(e_5-\mu_1) e_2 e_4 e_1 e_3 \cr
\bigg(\figbox{1.5cm}{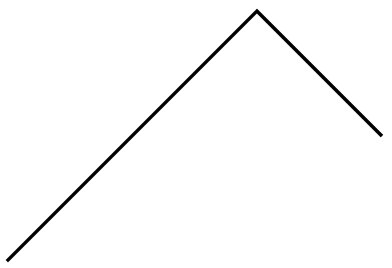}\bigg)_2~&=~\mu_1^3\figbox{1.5cm}{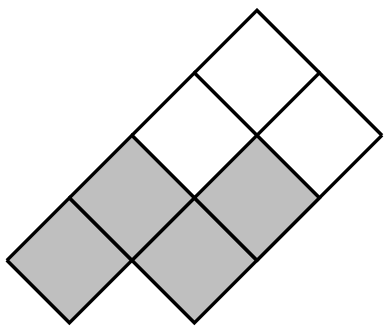}\cr
&=~ \mu_1^2\mu_2^{1/2}\mu_3^{1/2}
(e_4-\mu_2)(e_3-\mu_1)(e_5-\mu_1) e_2 e_4 e_1 e_3 \cr}}
where we have represented the grey and white boxes corresponding to each
walk diagram. Note e.g. for the last element of \sixtwo\ that
the rightmost white box is counted to have height $1$ (instead of 2)
because this height is the relative height w.r.t. the grey floor on the
same strip, which is already at height 1. 
This construction results in the following change of basis $1 \to 2$
\eqn\chbaot{ (a)_2~=~\sum_{f(a) \subset b \subset a} P_{b,a} \ (b)_1 }
with possibly non-vanishing matrix elements $P_{b,a}$ only for the 
walks $b \in W_n^{(h)}$ such that $b$ is above the floor of $a$
($f(a)\subset b$) and below $a$ ($b \subset a$).
Like in the meander case of Sect.3, we can arrange the walk diagrams by
growing length (number of boxes, grey and white), and make the 
matrix $P$ upper triangular.
\par
With this definition, the basis 2 satisfies the following
\par
\deb
\noindent{\bf PROPOSITION 3 :}
\par
The basis 2 elements are orthonormal with respect to
the bilinear form \bilin, namely
\eqn\proptwo{ \big( (a)_2,(b)_2\big)~=~ \delta_{a,b} 
\qquad {\rm for}\ {\rm all}\ 
a,b \in W_n^{(h)} }
\fin
\par
This result will be proved in the remainder of this section. 
Note first, in comparison with the meander case (proposition 1),
that no stronger statement (generalizing the lemma 1) will hold here
for the products of elements of ${\cal I}_n^{(h)}(q)$. This is because,
as mentioned earlier, ${\cal I}_n^{(h)}(q)$ is no longer
an ideal, hence we have no good control of what the product of two
elements of ${\cal I}_n^{(h)}(q)$ can be. Thus, instead of resorting
to the multiplication of elements, we will directly consider
the bilinear form \bilin. The main forthcoming
results (lemmas 2, 3 and 4 below)
will deal with reexpressions and simplifications of this bilinear form,
when evaluated on two elements of ${\cal I}_n^{(h)}(q)$. 
In particular, the lemma 3 will give a reexpression in terms of 
the form \bilin, evaluated respectively on elements of ${\cal
I}_{n-2p}^{(h)}(q)$ and ${\cal I}_{p}(q)$, which will enable
us to use the results of Sect.3, namely the proposition 1, to
eventually compute \proptwo.
\par
To prove the proposition 3, we need a few more definitions.
As we are basically dealing with elements
of the basis 2, it will be useful to trade
the usual notion of walk diagram $a\in W_n^{(h)}$ for that
of {\it bicolored box diagram}, namely the corresponding 
pictorial representation using grey and white box addition,
i.e. the arrangement of grey and white boxes forming $(a)_2$.
For convenience, we still denote by $(a)_2$ the bicolored
box diagram corresponding to $(a)_2$, with $a\in W_n^{(h)}$.
\fig{The bicolored box diagram corresponding to
an element $a\in W_{n}^{(n-10)}$ for all $n\geq 14$. 
The width of the diagram is $w=5$. It is decomposed into
5 strips $s_j(a)$, $j=1,2,...,5$.}{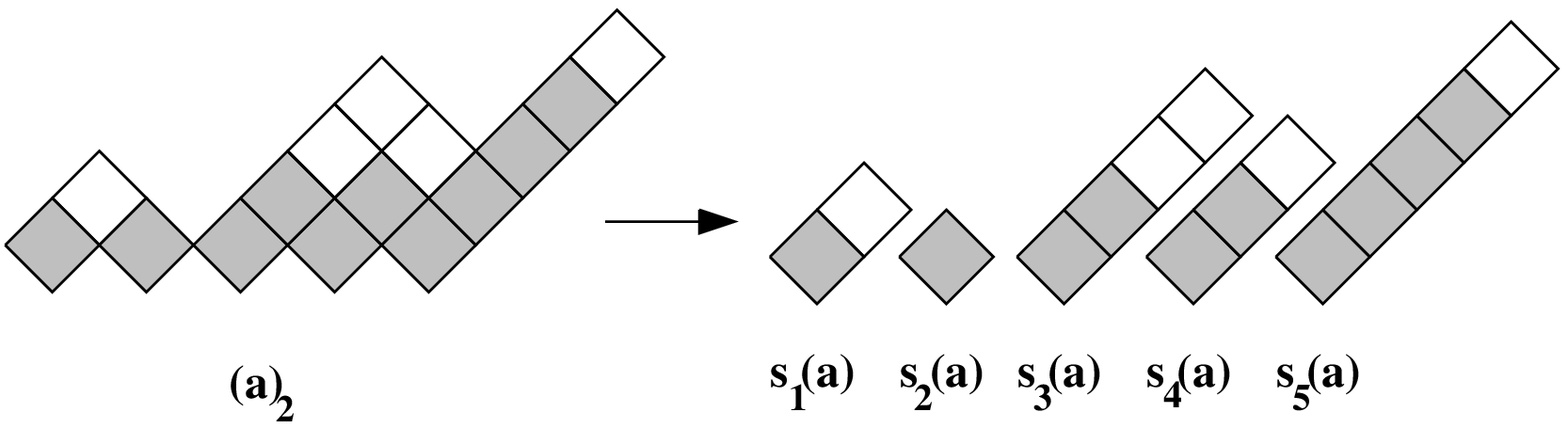}{9.cm}
\figlabel\bicostrip
Such a bicolored box diagram may be viewed as the succession
of {\it strips} $s_1(a)$, $s_2(a)$, ..., $s_w(a)$, made of
a succession of
grey, then white boxes of consecutive positions and heights.
The number $w$ stands for the number of these strips, namely
the {\it width} of the base of $(a)_2$, i.e. the number of grey boxes 
of height 0 in $(a)_2$. Note that for all $a\in W_n^{(h)}$,
the element $(a)_2$ has width $w=(n-h)/2$.
Moreover, we have the following identity between elements of 
${\cal I}_n^{(h)}(q)$
\eqn\idelem{ (a)_2~=~\mu_1^{n/2} s_1(a) s_2(a) ... s_w(a) }
by considering the strips (i.e. successions of grey and white boxes)
as elements of the Temperley-Lieb algebra.
\par
To proceed with the proof of proposition 3, we will
compute the quantity 
$\big( (a)_2,(b)_2\big)={\rm Tr}\big((a)_2^t (b)_2\big)$.
The strategy is the following. We will start by comparing
the rightmost
strips $s_w(a)$ and $s_w(b)$ of $(a)_2$ and $(b)_2$. 
Both are a succession of grey boxes, topped by one white box,
in the form
\eqn\forty{ s~=~\sqrt{\mu_{2}\over \mu_{1}} 
(e_{2w+j-2}-\mu_{1}) e_{2w+j-3} 
e_{2w+j-4} ... e_{2w} e_{2w-1}  }
with possibly different values of $j=j_a$ or $j_b$, 
the total size 
(total number of boxes) of the strip. 
Note that if $j_a=1$, $s_w(a)$ is reduced to a single grey
box, without white box on top (this is the case when $(a)_2$
only has one floor of height 0).
We have the first result 
\par
\deb
\noindent{\bf LEMMA 2 :}
\par
\noindent{}
For all $a,b\in W_n^{(h)}$, and $w=(n-h)/2$, if $s_w(a)\neq s_w(b)$,
then $\big( (a)_2,(b)_2\big)=0$.
\fin
\par
If $s_w(a)\neq s_w(b)$, then these strips have different size. 
Let us assume that $j_a<j_b$. Writing
$\big( (a)_2,(b)_2\big)={\rm Tr}\big((b)_2 (a)_2^t\big)$, and
$(b)_2=B s_w(b)$, $(a)_2=A s_w(a)$, we have
\eqn\weha{ \big( (a)_2, (b)_2\big)~=~{\rm Tr}\big( B s_w(b) 
s_w(a)^t A^t\big)}
In this expression, we now transfer the boxes of $s_w(b)$ onto $(a)_2^t$,
starting from the lowest one, up to the top of $s_w(b)$. These boxes now act
on $s_w(a)^t$ from below.
Thanks to the relation $e_i e_{i-1} e_i=e_i$, the first $j_a-2$ grey
boxes of the strip $s_w(a)^t$ are annihilated by the action of the first  
$j_a-1$ grey boxes of $s_w(b)$, namely
\eqn\dispol{\eqalign{ s_w(b) s_w(a)^t~&=~ 
{\mu_2 \over \mu_1}
(e_{2w+j_b-2}-\mu_{1})e_{2w+j_b-3}...e_{2w}e_{2w-1}^2 \cr
&\times 
e_{2w}...e_{2w+j_a-3} (e_{2w+j_a-2}-\mu_{1})\cr
&=~{\mu_2 \over  \mu_1^{2}} 
(e_{2w+j_b-2}-\mu_{1})e_{2w+j_b-3}...e_{2w+1}e_{2w}
e_{2w+1}\cr
&\times \ ...e_{2w+j_a-3} (e_{2w+j_a-2}-\mu_{1})\cr
&=~{\mu_2 \over  \mu_1^{2}} 
(e_{2w+j_b-2}-\mu_{1})e_{2w+j_b-3}...e_{2w+j_a-3}
(e_{2w+j_a-2}-\mu_{1})\cr }}
(Note that the last factor 
$(e_{2w+j_a-2} -\mu_1)$
must be replaced by $e_{2w+j_a-2}=e_{2w-1}$
in the case $j_a=1$, but this does not alter the following discussion.).
Let us now transfer in the same way all the boxes of $s_{w-1}(b)$,
$s_{w-2}(b)$, ..., $s_1(b)$ onto $(a)_2^t$. But these occupy
only positions $k\leq 2w+j_b-3$, and the largest position
$k=2w+j_b-3$ may only be occupied by a white box.
Hence, after the transfer of $(b)_2$ onto $(a)_2^t$
is complete, the resulting element is a linear combination of the form
\eqn\forlicop{\eqalign{
(b)_2 (a)_2^t~&=~ \alpha \  C' \ (e_{2w+j_b-2}-\mu_1)e_{2w+j_b-3} \  
C \cr
&+\beta \  D' \  e_{2w+j_b-3}(e_{2w+j_b-2}-\mu_1)e_{2w+j_b-3} \  
D \cr} }
where $C,D,C',D'$ are elements of the Temperley-Lieb algebra
only involving the generators $e_k$, $k<2w+j_b-3$, and $\alpha$ and
$\beta$ two complex coefficients, coming from the various normalization
factors. The second term in
\forlicop\ vanishes identically, thanks to the identity
$e_i (e_{i+1}-\mu_1) e_i=0$.  We are therefore left with
\eqn\lefwiop{\eqalign{ \big( (a)_2, (b)_2\big)~&=~\alpha \ {\rm Tr}\big( 
C'(e_{2w+j_b-2}-\mu_1) e_{2w+j_b-3} C\big) \cr
&=~\alpha \  {\rm Tr}\big( 
(e_{2w+j_b-2}-\mu_1) e_{2w+j_b-3} C C'\big) \cr}}
To show that this expression vanishes, let us use the string
representation of the $e_i$, and the definition of the trace as
computing $q^L$, where $L$ is the number of loops of the string
representation of the element, after identification of the upper
and lower ends of its strings.
In this picture (setting $i=2w+j_b-2$, $CC'=E$, and taking the
adjoint of the expression
in the trace, which does not change its value), we have 
\eqn\picalc{\eqalign{
{\rm Tr}\big(E(e_1,e_2,...,e_{i-2}) &e_{i-1}(e_i-\mu_1)\big) \cr
&=~\figbox{3.5cm}{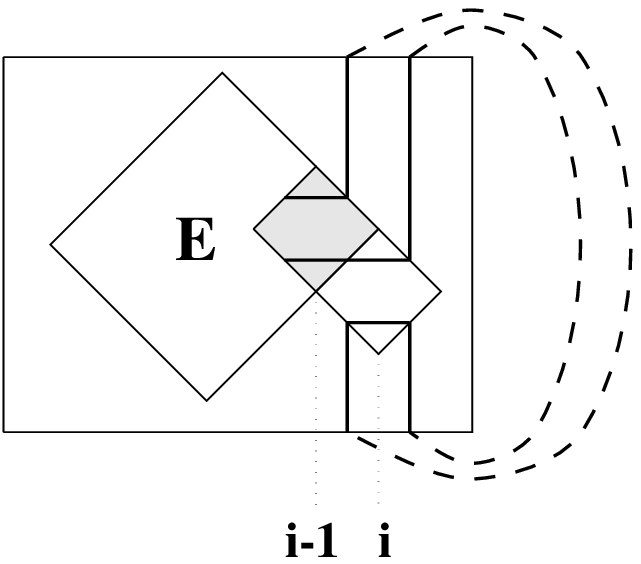}-\mu_1\figbox{3.5cm}{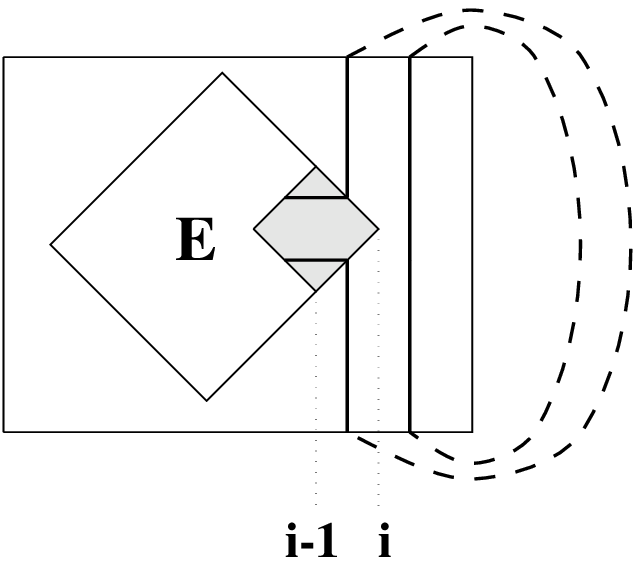} \cr
&=~\sum_r \alpha_r ( q^L -\mu_1 q^{L+1}) \cr
&=~0\cr}}
where we have expanded $E(e_1,e_2,...,e_{i-2})$ as a linear combination
of diagrams involving only grey boxes with positions $k\leq i-2$.
In each of these diagrams, the second term has always one more loop than
the first one, hence the cancellation, with the factor $\mu_1=q^{-1}$.
This completes the proof of the lemma 2.
\par
The lemma 2 guarantees that $\big( (a)_2,(b)_2\big)=0$ as soon as
the last strips $s_w(a)$ and $s_w(b)$ are distinct. In the latter
case, proposition 3 is therefore proved.
Let us assume now that $(a)_2$ and $(b)_2$ have the same last strip,
say with $j$ boxes.
Then both $a$ and $b$ have a rightmost floor of height $H=j-1$. 
Let $p_a$ and $p_b$ denote their respective widths, namely
the respective numbers
of grey boxes of height $j-1$ forming this floor in $a$ and $b$.
Two situations may occur for these floors 
\item{(i)} they have
the same width $p_a=p_b$. In this case, we will show that
$\big( (a)_2,(b)_2\big)$ is factored into the bilinear
form \bilin\ evaluated on smaller diagrams, obtained by 
cutting $(a)_2$ and $(b)_2$ into two pieces (lemma 3 below).
\item{(ii)} the width of the rightmost
floor of $a$ is strictly smaller that that of the rightmost floor
of $b$ $p_a<p_b$. In this case, we will show that 
$\big((a)_2,(b)_2\big)=0$ (lemma 4 below).
\par
\noindent{}Let us treat these cases separately.
\par
\noindent{\bf CASE (i) :} the two rightmost floors of
$a$ and $b$ have the same width $p_a=p_b=p$.
We will simply {\it grind} 
the $j-2$ consecutive layers of grey boxes
underlying the floor of height $j-1$, and detach the corresponding
portions of $a$ and $b$, so that the quantity $\big( (a)_2,(b)_2\big)$
will factorize into a product of analogous terms, for smaller
diagrams (see lemma 3 below). 
More precisely, let us compute 
the quantity 
\eqn\qtiti{S(a) S(b)^t~=~ 
s_{w-p+1}(a) s_{w-p+2}(a)... s_w(a) s_w(b)^t...
s_{w-p+1}(b)^t }
involved in the computation of $(a)_2 (b)_2^t$. In $S$
of \qtiti, all the
strips involved have a floor of height $j-1$, i.e. have the form
\eqn\forinqti{\eqalign{
s_{w-m+1}~&=~ \diamond \diamond ...  \diamond 
e_{2w-2m+j-1}e_{2w-2m+j-2} ... e_{2w-2m+1}\cr 
&=~ {\tilde s}_{w-m+1} e_{2w-2m+j-2} ... e_{2w-2m+1}\cr } }
where each strip 
$\tilde s$ has a floor of only one grey box, topped
by white boxes.
The idea is to transfer the grey boxes from $s_w(a)$ to $s_w(b)$,
{\it from below}, just like we did in \dispol, and do it again 
for $s_{w-1}(a)$ and $s_{w-1}(b)$, etc... until we are left
only with the amputated strips $\tilde s$.
The final result simply reads
\eqn\finSres{ S(a) S(b)^t~=~{\tilde S}(a) {\tilde S}(b)^t~=~
{\tilde s}_{w-p+1}(a) {\tilde s}_{w-p+2}(a)... {\tilde s}_w(a) 
{\tilde s}_w(b)^t ... {\tilde s}_{w-p+1}(b)^t }
This result implies the following
\par
\deb
\noindent{\bf LEMMA 3 :}
\par
If $a$ and $b\in W_n^{(h)}$ have identical 
rightmost floors of width $p$, then 
\eqn\letroi{ \big( (a)_2, (b)_2\big) ~=~ \big( (a')_2,(b')_2\big)\
\big( (a'')_2, (b'')_2\big) }
where 
\eqn\whereaap{\eqalign{
(a')_2~&=~\mu_1^{n-2p \over 2} s_1(a) s_2(a) ... s_{w-p}(a) \cr
(b')_2~&=~\mu_1^{n-2p \over 2}s_1(b) s_2(b) ... s_{w-p}(b) \cr
(a'')_2~&=~\mu_1^p {\tilde s}_{w-p+1}(a) ... {\tilde s}_{w}(a)\cr
(b'')_2~&=~\mu_1^p {\tilde s}_{w-p+1}(b) ... {\tilde s}_{w}(b)\cr}}
\fin
\par
The normalizations in \whereaap\ are chosen to guarantee that
all the
elements $(a')_2$, $(a'')_2$, ... have norm 1, as we will see below.
The lemma 3 will follow from the application of \finSres\ to the 
computation of $\big( (a)_2,(b)_2\big)={\rm Tr}\big( (a)_2 (b)_2^t\big)$,
Indeed, we simply write
\eqn\sipwri{\eqalign{ 
\big( (a)_2, (b)_2\big)~&=~\mu_1^{2p}\  {\rm Tr}\big(
(a')_2 S(a) S(b)^t (b')_2^t\big)\cr
&=~\mu_1^{2p}\ 
{\rm Tr}\big( (a')_2 {\tilde S} {\tilde S}^t (b')_2^t\big)\cr 
&=~{\rm Tr}\big( (a'')_2 (b'')_2^t (b')_2^t (a')_2 \big) \cr
&=~\big( (a'')_2,(b'')_2\big) \times \big( (a')_2,(b')_2\big) \cr}}
In the last step, we have noted that $(a'')_2 (b'')_2^t$ involves
only generators $e_k$ with positions $k\geq 2w-2p+j-1$ (position
of the leftmost grey box in ${\tilde S}(a)=(a'')_2$), whereas
$(a')_2 (b')_2^t$ involves only generators $e_k$ with
$k \leq 2w-2p+j-3$ (maximum
position of the rightmost (white) box in $(a')_2$).
The last line of \sipwri\ 
follows then from the locality of the trace, namely that 
for any two sets of positions $I$, $J$, with $i<j-1$
for all $i\in I$, $j\in J$
\eqn\loctra{ {\rm Tr}(\prod_{i\in I} e_i 
\prod_{j\in J} e_j)= {\rm Tr}(\prod_{i\in I} e_i)\ {\rm Tr}(
\prod_{j\in J} e_j) }
which follows from the definition of the trace (the loops arising
from the two terms are independent).
\par
In \letroi, the bilinear forms 
$\big((a)_2,(b)_2\big)$,
$\big((a')_2,(b')_2\big)$ and
$\big((a'')_2,(b'')_2\big)$,
are respectively evaluated in
the spaces ${\cal I}_n^{(h)}(q)$, ${\cal I}_{n-2p}^{(h)}(q)$ and
${\cal I}_{2p+1}^{(1)}(q)$.
Let us concentrate on the last term $\big( (a'')_2,(b'')_2\big)$.
There is a simple morphism $\varphi$ of algebras between
${\cal I}_{2p+1}^{(1)}(q)$ and 
${\cal I}_{2p+2}^{(0)}(q)={\cal I}_{p}(q)$
\eqn\morph{ \varphi(E)~=~
\sqrt{\mu_1} E\ e_{2p+1} \in {\cal I}_{p}(q) \qquad 
\forall \ E \in {\cal I}_{2p+1}^{(1)}(q)}
The morphism $\varphi$ consists simply in adding the missing
rightmost grey box ($e_{2p+1}$) to complete the floor of
$E$ into that of the ideal ${\cal I}_{p}(q)$. 
Moreover, 
we have added an ad-hoc  multiplicative normalization factor $\mu_1$. 
With this normalization, we have the following
simple correspondence between traces over the two spaces
\eqn\tracores{ \eqalign{
{\rm Tr}\big( \varphi(E) \varphi(F)^t\big)~&=~ 
\mu_1\mu_1^{-1} \ {\rm Tr}\big( E \, e_{2p+1} \, F^t\big) \cr 
&=~\figbox{3.cm}{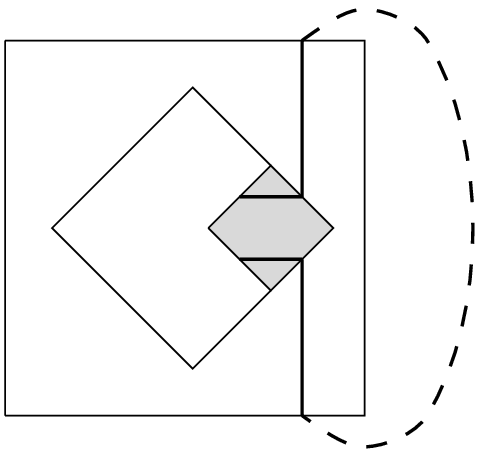}~
=~\figbox{2.4 cm}{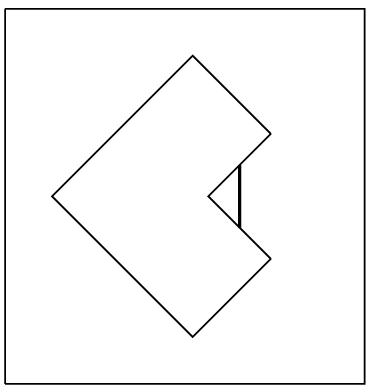} \cr
&=~ {\rm Tr}\big ( E F^t \big)\cr}}
where we have used $e_{2p+1}^2=\mu_1^{-1} e_{2p+1}$, then
transferred all the boxes of $F^t$ onto $E$, and represented
the result in the pictorial
string-domino representation of the trace (see Fig.\dostri), 
to show that the presence of the 
grey box does not change the value of the trace (it does not affect
the structure of the loops).  
Using this fact, we can now apply the result of the proposition 1
of Sect.3.3 above to the factor $\big( (a'')_2,(b'')_2\big)$
of \letroi, by simply interpreting
$\varphi\big( (a'')_2\big)$ and $\varphi\big( (b'')_2\big)$
as elements of ${\cal I}_{p}(q)$. 
We conclude that
\eqn\concluio{\eqalign{ 
\big( (a'')_2,(b'')_2\big)~&=~ \bigg(
\varphi\big( (a'')_2\big), \varphi\big( (b'')_2\big) \bigg) \cr
&=~ \delta_{a'',b''} \cr }}
Hence, if $a''\neq b''$, the bilinear form vanishes, and the proposition
3 follows. If $a''=b''$, we go back to the 
beginning of our study, with now $(a)_2$ and $(b)_2$ replaced
with $(a')_2$ and $(b')_2 \in {\cal I}_{n-2p}^{(h)}(q)$. If only the
case (i) occurs, we will dispose successively of each portion
of $a$ and $b$ above their common successive floors (as above), and
get an expression
\eqn\exofab{ \big( (a)_2, (b)_2\big) ~=~ \prod_{
{\rm portions}\ a'',b'' \atop
{\rm above} \ {\rm successive}\ {\rm floors}} \delta_{a'', b''} }
If the case (ii) occurs, the result will vanish, as we will see now.
\par
\noindent{\bf CASE (ii) :} The diagrams $(a)_2$ and $(b)_2$
have a rightmost floor, of same  height $j-1$, but with different
widths $p=p_a<p_b$. 
As in the case (i), we concentrate on the portions
of $a$ and $b$ above this rightmost floor, over a width $p$,
namely consider
\eqn\conS{\eqalign{ 
S(a)~&=~ s_{w-p+1}(a) s_{w-p+2}(a) ... s_{w}(a)\cr
S(b)~&=~ s_{w-p+1}(b) s_{w-p+2}(b) ... s_w(b) \cr}}
To compute the quantity $\big( (a)_2,(b)_2\big)$, we now write
$(a)_2=(a')_2 S(a)$ and $(b)_2=(b')_2 S(b)$, and
get
\eqn\getrac{\eqalign{ \big( (a)_2, (b)_2 \big) ~&=~
{\rm Tr} \big( (a')_2 S(a) S(b)^t (b')_2^t\big) \cr
&=~{\rm Tr} \big( (b')_2^t (a')_2 {\tilde S}(a)
{\tilde S}(b)^t \big)\cr}} 
Using the cyclicity of the trace and the symmetry of \bilin, 
we may also write
\eqn\maywrt{\eqalign{
\big( (a)_2,(b)_2\big)~&=~{\rm Tr}
\big( {\tilde S}(a)^t (a')_2^t (b')_2 {\tilde S}(b)\big)\cr
&=~{\rm Tr} \big( (a')_2^t {\tilde S}(a)^t(b')_2 {\tilde S}(b)\big)\cr}}
We have used the commutation of ${\tilde S}(a)$, which involves
only boxes of positions $\geq \alpha=2w-2p+j-1$, with $(a')_2$,
which only involves boxes of positions $\leq \alpha-2$ (as its rightmost
floor has now an height $<j-1$).
Let us now compute \maywrt\ by transferring the {\it white boxes}
of ${\tilde S}(a)^t$ onto $(b')_2 {\tilde S}(b)$.
Once this transfer is complete, $(b')_2 {\tilde S}(b)$ is replaced by
a linear combination of diagrams $(c')_2 {\tilde S}(c)$
with all possible box
additions/subtractions induced by the process of transfer. 
Note that the left portion $(b')_2$ of $(b)_2$ is also affected,
as these boxes act on both $(b')_2$ and ${\tilde S}(b)$.  
For notational simplicity, we have
used the denomination $c'$ for the left
part of $c$, so that we still have $(c)_2=(c')_2 S(c)$.
We are
then left with the transfer of the first layer of grey boxes of 
${\tilde S}(a)^t$,
namely those of height $j-1$.  To get a non-zero result, those must
only hit minima or maxima on $(c')_2 {\tilde S}(c)$
(the action of a grey box on a white slope vanishes, according
to \greybw\-\nediagr). 
Concentrating on the 
configuration of $(c')_2 {\tilde S}(c)$ above
the position $\alpha=2w-2p+j-1$ (namely
the configuration of $(c')_2$ above the leftmost grey box in 
${\tilde S}(c)$), only two situations may yield a non-zero answer 
\item{(a)} $(c')_2$ has no white box above the position $\alpha$.
\item{(b)} $(c')_2$ has a white maximum at $\alpha$. In this case,
this maximum is necessarily at height $j+2$ 
(white box of height $j+1$, hence of relative height $2$ w.r.t. the floor), 
because no white slope is allowed at
any of the positions $\alpha$, $\alpha+2$, ..., $\alpha+2p-2=2w+j-3$,
and the rightmost white box of ${\tilde S}(c)$ has an height $\leq j$,
hence a relative height $\leq 1$. 
Let us transfer this white box {\it back} onto what is left of
${\tilde S}(a)^t$ (call it ${\tilde S}(d)^t=e_\alpha e_{\alpha+2}
...e_{\alpha+2p-2}$). 
Actually, this diagram has now a grey maximum at the position
$\alpha$ (this is the position of the leftmost grey box in the floor of
$S(a)$). The white box acts on this grey maximum as
$\sqrt{\mu_3/\mu_2}(e_\alpha-\mu_2)e_\alpha=e_\alpha/\sqrt{\mu_2 \mu_3}$,
hence is eliminated up to some multiplicative constant. We therefore
end up in a situation where $(c')_2\to (c'-\diamond_\alpha)_2$ 
has no white box above the position
$\alpha$ hence in the case (a) above.
\par
\noindent{} In either case, we end up in a situation where $\alpha$
is a floor-end on both diagrams $(a')_2 {\tilde S}(d)$ and 
$(c'')_2 {\tilde S}(c)$, where $c''=c'$ in the case (a) and 
$c''=c'-\diamond_\alpha$ in the
case (b). 
So we can reexpress
\eqn\rexdefr{\eqalign{ 
\big( (a)_2,(b)_2\big)~&=~\sum_c \lambda_c {\rm Tr}\big(
(a')_2^t (c'')_2 {\tilde S}(c) {\tilde S}(d)^t\big) \cr
&=~\sum_c \lambda_c  
\figbox{4.cm}{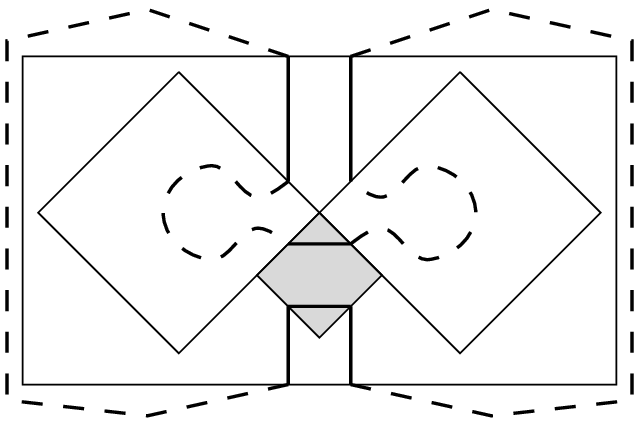}\cr
&=~ \mu_1 \sum_c \lambda_c \figbox{4.cm}{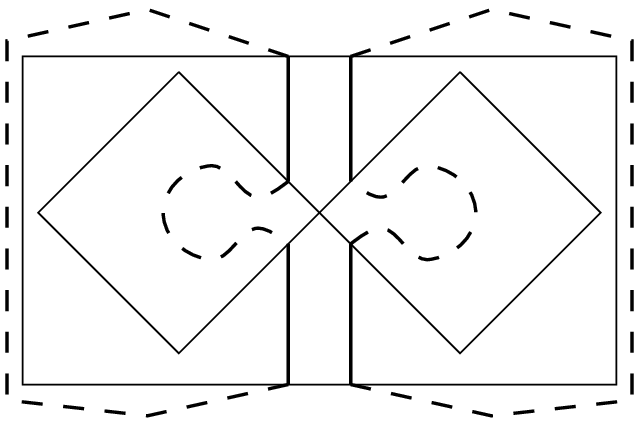} \cr
&=~\mu_1 \sum_c \lambda_c 
\big( (a')_2,(c'')_2\big) \ {\rm Tr}\big({\tilde S}(c)
{\tilde S}(d)^t\big) \cr}}
where, by using the string-domino picture, we have removed
the grey box linking the left and right parts of the operator
in the trace,
at the expense of creating a new loop, hence the 
extra factor of $\mu_1=q^{-1}$.
Note that in this argument it was crucial that there should be no
white box above the grey box we have removed, further linking the
left and right parts: this is why we had to go through the case (b)
above and modify $c'\to c''=c'-\diamond$ 
to get back to the situation (a).
\par
Now the main feature of $(c'')_2$ is that it has still a rightmost grey
floor of height $j-1$, whereas by definition $(a')_2$
has a rightmost grey floor of height $<j-1$. Hence the rightmost strips
in both diagrams are distinct: $s_{w-p}(a')\neq s_{w-p}(c'')$. 
We can therefore apply the lemma 2, to conclude that 
\eqn\bilopr{ \big( (a)_2,(c'')_2\big) ~=~0}
in \rexdefr,
so that finally  $\big( (a)_2,(b)_2\big)=0$.
Hence we deduce the 
\par
\deb
\noindent{\bf LEMMA 4:} 
\par
For any two bicolored box diagrams $(a)_2$ and $(b)_2$, with 
rightmost floors of same height $j-1$, but of differents widths
$p_a<p_b$, we have
\eqn\lemfour{ \big( (a)_2,(b)_2\big)~=~ 0}
\fin
\par
The proof of the proposition 3 is now straightforward. 
We start with the two bicolored box diagrams $(a)_2$ and $(b)_2$.
If their rightmost strips are distinct, then $\big(
(a)_2,(b)_2\big)=0$ by the lemma 2. Otherwise, 
we focus our attention to their rightmost floors, which have
the same height $j-1$. If they have different widths, 
the lemma 4 above implies that $\big( (a)_2,(b)_2\big)=0$.
If they have the same width, the lemma 3 expresses
$\big( (a)_2,(b)_2 \big)=\prod \delta_{a'',b''}$, hence we
finally get that $\big( (a)_2,(b)_2\big)=0$ unless
$a$ and $b$ are identical, in which case
$\big( (a)_2,(a)_2\big)=1$. This completes the proof
of the proposition 3.
\par
\subsec{The semi-meander determinant: a preliminary formula} 
The semi-meander determinant \relsdetg\ follows
from the Gram determinant of the basis 1. The latter is best
expressed through the change of basis 1 $\to$ 2, in which the Gram
matrix is sent to the $c_{n,h}\times c_{n,h}$ identity matrix $I$.
With the upper triangular matrix $P$ defined in \chbaot, this reads
\eqn\triangul{ P \Gamma_{n}^{(h)}(q) P^t~=~I}
Hence $\det \Gamma_n^{(h)}(q)=(\det P)^{-2}$.
The diagonal elements of $P$ are linked by the recursion relation
\eqn\recpel{ P_{a+\diamond_{i,\ell},a+\diamond_{i,\ell}}~=~
\sqrt{\mu_{\ell+1} \over \mu_\ell} P_{a,a} }
where the box addition $\diamond_{i,\ell}$ is performed at the
point $i$, and at {\it relative} height $\ell$, with respect to the
grey box floor in $a$. With the initial condition
$P_{a_n^{(h)},a_n^{(h)}}=\mu_1^{n/2}$ \basd\ for the 
fundamental walk diagram of $W_n^{(h)}$, this gives
\eqn\calpof{ P_{a,a}^2~=~ \mu_1^n
\prod_{{\rm white}\ {\rm boxes}\atop
{\rm of}\ a} {\mu_{\ell+1}\over \mu_\ell} }
where $\ell$ denotes the height of the white box addition,
relative to the grey floor in $a$.
\fig{The strips of white boxes on a walk $a\in W_n^{(h)}$
with $n=20$ and $h=6$. The walk is represented in a thick
black line. We have also represented the floor of grey boxes for this
walk. We have $(n-h)/2=7$ strips of white boxes, of respective
lengths $2$, $1$, $3$, $2$, $2$, $2$, $2$.}{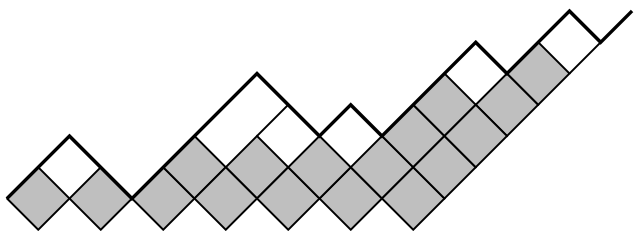}{7.cm}
\figlabel\bstrips
Like in the meander case, let us arrange the white boxes of any
bicolored box diagram corresponding to an
$a\in W_{n}^{(h)}$ into strips of white boxes, namely
sequences of white boxes with consecutive positions and heights,
added on top of the grey floor of $a$ 
(see Fig.\bstrips\ for an illustration). There are exactly $(n-h)/2$
such white strips. The strip length is now defined as the relative 
height of the
top of the
white box sitting on top of the strip (hence an empty strip has
length $1$). 
With this definition, we simply have
\eqn\defrop{ P_{a,a}^2~=~ \mu_1^{n+h\over 2}
\prod_{{\rm white} \ {\rm strips} \ {\rm of} \ a}
\mu_{\ell} }
where, in the product over the $(n-h)/2$ strips of $a\in W_n^{(h)}$,
$\ell$ stands for the strip length (all denominators have
been cancelled along the strips, except for the $\mu_1$ ones, which
have rebuilt the prefactor).
This yields the determinant of the basis 1
\eqn\graonrgt{ \det \Gamma_n^{(h)}(q)~=~
\prod_{a \in W_n^{(h)}}
P_{a,a}^{-2}~=~ 
\mu_1^{-(n+h)c_{n,h}/2}
\prod_{{\rm white}\ {\rm strips}\ {\rm of}\atop
{\rm all}\ a \in W_n^{(h)}} \mu_\ell^{-1} }
and thanks to \relsdetg, the semi-meander determinant
\eqn\sdefrtgy{ \det {\cal G}_n^{(h)}(q)~=~ 
\mu_1^{-h c_{n,h}} \prod_{{\rm white}\ {\rm strips}
\ {\rm of} \atop {\rm all}\  
a \in W_n^{(h)}} \mu_\ell^{-1} }
The latter can be recast into
\eqn\recsdet{ \det {\cal G}_n^{(h)}(q)~=~
\mu_1^{-h c_{n,h}}
\prod_{m=1}^{{n-h\over2}+1}
\big[ \mu_m \big]^{- s_{n,m}^{(h)} }}
where $s_{n,m}^{(h)}$ denotes the total number of white
strips of
length $m$ in all the bicolored box diagrams
corresponding to the walk diagrams
of $W_n^{(h)}$ (the notation is such that $s_{2n,m}^{(0)}=s_{2n,m}$
\concluop). Note also that 
the strips have all length $\leq (n-h)/2+1$, hence the upper bound in
the product in \recsdet. 
The formula of theorem 2 will follow from the explicit 
computation of the numbers $s_{n,m}^{(h)}$. 
\par
This will be done in two steps. 
The first step (Proposition 4, Sect.4.4 below)
consists in arranging the $s_{n,m}^{(h)}$ walks above
according to their floor configuration (namely their configuration
of grey boxes). 
The second step (Proposition 5, Sect.4.5 below) consists in enumerating the 
walks with minimal floor configurations (namely made
of only one layer of grey boxes). 
Finally in the proposition 6, Sect.4.6, 
the combination of these two results
will eventually lead to a formula for $s_{n,m}^{(h)}$, 
which will complete the proof of the theorem 2.
\par
\subsec{Enumeration of the floor configurations}
In this first step, we note that
many different diagrams $a\in W_n^{(h)}$ have the same contribution
to \sdefrtgy, namely those with identical white strips, but different
floors of grey boxes.  Assembling all these contributions leads
to the following formula for $s_{n,m}^{(h)}$
\par
\deb
\noindent{\bf PROPOSITION 4 :}
\par
\eqn\profour{ s_{n,m}^{(h)}~=~ \sum_{k\geq 0} {h+k-1\choose k} 
f_{n-h+2,m,k} }
where $f_{2n,m,k}$ denotes the total number of walk diagrams 
$a\in W_{2n}$ with 
$k+1$ floors of height 0, and with a marked top of
strip of length $m$.
\fin
\par
\fig{A typical walk diagram $a\in W_n^{(h)}$ is represented in thick
black line on the upper
diagram. 
We have also represented its floor of grey boxes, and the
white boxes topping it.
The floor of grey boxes in $a$
is a succession of a number $k+1$ of horizontal floors, 
$F_0$, $F_1$, ..., $F_k$,
with respective heights $H_0=0$, $H_1$, $H_2$, ..., $H_k\geq 0$.
The conjugates of $a$ are obtained by varying these heights,
without changing the white strips of $a$ (this is done
by letting the floors slide along the dashed lines separating them). 
The minimal conjugate $\hat a\in W_{n-h+2}$ of $a$
is represented below it: 
it has $H_1=H_2=...=H_k=0$. The floor-ends are indicated by 
arrows.}{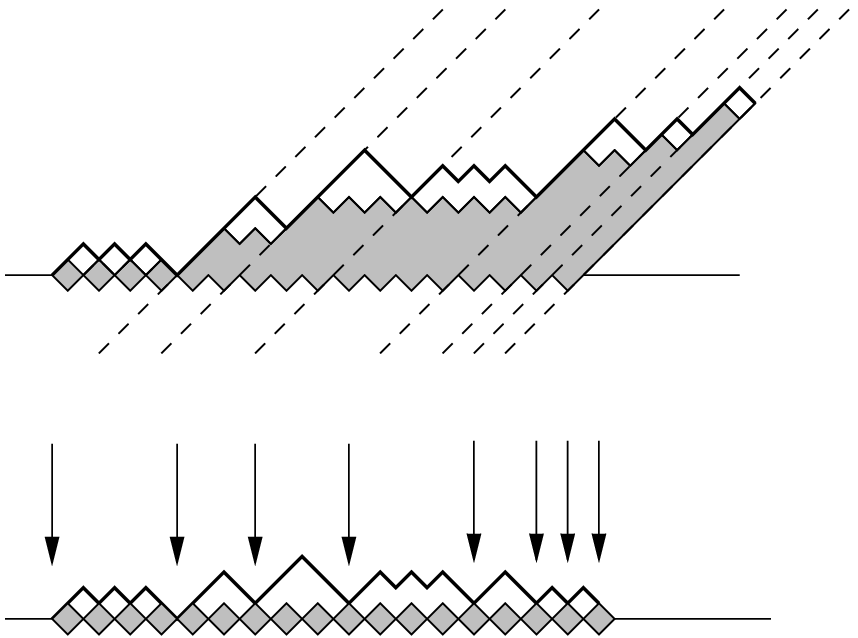}{7.cm}
\figlabel\floors
Indeed, as illustrated in Fig.\floors, 
in any walk diagram $a\in W_n^{(h)}$, the floor of grey boxes
may be viewed as a succession of a number, say $k+1$ of consecutive
horizontal floors of grey boxes $F_0$, $F_1$, ..., $F_k$, with respective 
heights $H_0=0$, $H_1$, ..., $H_k$, and $H_j\geq 0$ for all $j\geq 1$. 
The leftmost floor $F_0$, of height $H_0=0$, is made of one layer of
grey boxes of the form $e_1 e_3 e_5 ...$, and occupies a segment
$I_0=\{ i_0=0,1,2,...,i_1-1\}$ of positions
(we include here the case when $F_0=\emptyset$, i.e.
$I_0=\{0\}$, corresponding to the
case $J_0\neq \{0\}$ of \jdef). It is topped by white
strips of arbitrary lengths.
Any horizontal floor $F_j$, $j=1,2,...,k$, of height $H_j\geq 0$,
is a parallelogram made of $H_j+1$ horizontal
layers of grey boxes, whose base occupies a segment of positions 
$I_j=\{i_j,i_j+1,i_j+2,...,i_{j+1}\}$, with $i_{k+1}=n-h+1$.
What distinguishes these floors from $F_0$ is that they are necessarily
topped with {\it at least} two layers of white boxes, resulting
in white strips of lengths $m\geq 2$ only.
The separation between two consecutive floors of height $H_j\geq 0$ 
is formed by the strips
of length 2 (with one white box), as illustrated in
the upper diagram of Fig.\floors,
where the floor separations are indicated by dashed lines. 
The various floor heights are subject to the constraint
\eqn\consflor{ 0 \leq H_1 \leq H_2\leq ...\leq H_k \leq h-1 }
arising from the original definition of the floor of a 
walk $a \in W_n^{(h)}$ (the floor $F_j$ is always 
of lesser or equal height than $F_{j+1}$).
\par
By varying only the heights $H_1$, $H_2$, ..., $H_k$ subject to 
\consflor, and by keeping the white strips fixed,
we describe the set of all {\it conjugates} of a given 
walk diagram $a\in W_n^{(h)}$. There are therefore
\eqn\conjnum{ |\{(H_1,...,H_k)\in \IN \ {\rm s.t.} \ 
0\leq H_1 \leq H_2 \leq...\leq H_k\leq h-1\}|~=~ {h+k-1 \choose k}} 
such conjugates for each diagram $a \in W_n^{(h)}$ with $k+1$ floors.
We now choose among the conjugates of $a$, 
the {\it minimal} one,
namely that with $H_1=H_2=...=H_k=0$, which we denote by $\hat a$
(the bottom diagram of Fig.\floors).
We may amputate this diagram from the final slope with positions
$n-h+2$, $n-h+3$, ..., $n$, and view it as a diagram
$\hat a\in W_{n-h+2}$. Indeed, the diagram $\hat a$ has $h(n-h+1)=1$, 
the height of the rightmost floor-end, hence we may complete it by 
$h(n-h+2)=0$ into an element of $W_{n-h+2}$.
Denoting by $f_{2n,k,m}$ the total number of walks of $W_{2n}$ with 
$k+1$ floors of height 0, and
with a marked top of strip of length $m$, the proposition 4
follows, by enumerating these $f_{n-h+2,k,m}$ walk diagrams
with a marked top of strip of length $m$, and weighing each of them by
the number of its conjugates \conjnum.
\par
\subsec{The mapping of walk diagrams}
The second step of the calculation of $s_{n,m}^{(h)}$ is the 
computation of the numbers $f_{2n,k,m}$ appearing in \profour.
The result reads
\par
\deb
\noindent{\bf PROPOSITION 5 :}
\par
The total number $f_{2n+2,k,m}$ of walks in $W_{2n+2}$, with $k+1$ floors
of height 0, and with a marked top
of strip of length $m$ reads
\eqn\numread{
\eqalign{ f_{2n+2,k,m}~&=~c_{2n-k,2m+k}+ k c_{2n-k,2m+k-4}
\qquad {\rm for}\ m \geq 2, \ n \geq 1 \cr
f_{2n+2,k,1}~&=~ c_{2n-k,k+2} \qquad {\rm for} \ m=1, \ 
n \geq 1 \cr}  }
where the numbers $c_{n,h}$ are defined in \oparnum.
Here we have excluded the trivial case $n=0$, for which no strip 
appears, hence 
\eqn\fzer{f_{2,k,m}~=~0 \qquad  {\rm for} \ {\rm all}\  k \
{\rm and}\  m}
\fin
\par
To prove this proposition, we will construct a bijection from the set
of walk diagrams of $W_{2n+2}$ with $k+1$ floors of height 0,
and with a marked top of strip of
length $m\geq 2$ to (i) the set $W_{2n-k}^{(2m+k)}$  (ii) $k$ copies
of the set $W_{2n-k}^{(2m+k-4)}$, which will prove \numread,
as $|W_n^{(h)}|=c_{n,h}$. (In the case $m=1$, only the part (i) will
apply, namely, we will construct a bijection between the walks of
$W_{2n+2}$ with a marked end of empty strip (length 1) and
$W_{2n-k}^{(k+2)}$.). 
\par
We start from $a\in W_{2n+2}$, with $k+1$ floors, all of height 0, 
and with a marked top of strip of length $m$.
Two cases may occur: 
\item{(i)} The marked top of strip lies above the 
leftmost floor ($F_0$).
In this case, we will 
construct a walk $b\in W_{2n-k}^{(2m+k)}$
by a cutting-reflecting-pasting procedure on $a$, analogous to that
used in the meander case. This will produce the first term in \numread.
\par
\item{(ii)} The marked top of strip lies above one of the $k$ other 
floors ($F_1$, $F_2$, ..., $F_k$). This is possible only if $m\geq 2$,
as there is no empty strip above these floors, by definition. 
By a circular permutation of the $k$ floors, we
can always bring the block containing the marked point to the right.
We therefore have a $k$-to-one mapping to the situation where
the marked strip is above the rightmost floor. 
This $k$-fold circular permutation
symmetry is responsible for the factor $k$ in 
the second term of \numread.
The diagrams with the marked top of strip of length $m$
above the rightmost ($F_k$) floor
are then
mapped to the walk diagrams with a marked top of strip at height $m-2$ 
above the leftmost ($F_0$) floor considered in the case (i), hence
to the set $W_{2n-k}^{(2(m-2)+k)}=W_{2n-k}^{(2m+k-4)}$ (the case $m=2$
will have to be treated separately).
Together with the multiplicity factor $k$ this will produce the 
second term of \numread.
\par
Let us now construct the maps for the cases (i) and (ii) above.
\par
\noindent{\bf CASE (i) :}
We start from a walk diagram $a\in W_{2n+2}$, 
with $k+1$ floors of height 0, 
and with a marked top of strip
of length $m$ above its leftmost floor $F_0$, say at position $i$.
The point $(i,h(i)=m)$ separates $a$ into a left $L$ and right $R$
parts, respectively such that $L\in W_i^{(m)}$ and 
$\bar R\in W_{2n+2-i}^{(m)}$. Reflecting $L$ and pasting it again at the 
left end of $R$, we create a walk diagram $b'$, whose 
reflection $\bar{b'}\in W_{2n+2}^{(2m)}$.
To construct the eventual image $b\in W_{2n-k}^{(2m+k)}$ of $a$, 
we  perform the
following amputations of the walk $b'$. We will
suppress some pieces of $b'$ 
at each separation of floor, 
according to the following rules
\eqn\rulamput{\eqalign{ 
(1)\ \ \figbox{2.2cm}{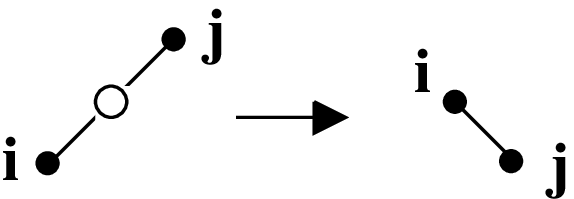} &\ (h\to h+3, o\to o-1) \cr
(2)\ \ \figbox{2.2cm}{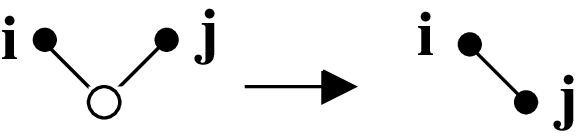} &\ (h\to h+1, o\to o-1) \cr
(3)\ \ \figbox{2.2cm}{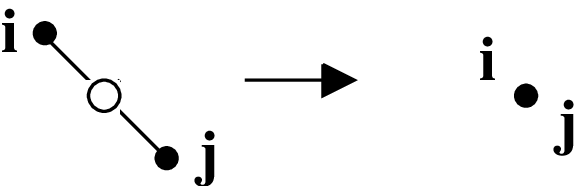} &\ (h\to h-2, o\to o-2) \cr }}
where we have represented the floor end by an empty circle,
and where we indicate the change in final height ($h$) and in the
order ($o$) resulting from the amputation.
Considering that the rules in \rulamput\ apply respectively 
(1) to the first
floor separation only (between $F_0$ and $F_1$), 
(2) to the $k-1$
intermediate floor separations 
(between $F_j$ and $F_{j+1}$, $j=1,2,...,k-1$), 
and (3) to the
rightmost floor end (right end of $F_k$), 
we get an overall change from the initial values
$(h=2m, o=2n+2)$ of the height and order of $b'$ to the
amputed $b''$ with
\eqn\chanoh{ h\to h+3+(k-1)-2=2m+k \qquad o\to o-1-(k-1)-2=2n-k }
Hence taking $b={\bar {b''}}$, we get an element of $W_{2n-k}^{(2m+k)}$.
\par
To prove that this mapping is bijective, let us compute its inverse.
Starting from $b\in W_{2n-k}^{(2m+k)}$, let $i$ be the position
of the {\it rightmost} intersection between $b$ and the line $h=m+k$
at an ascending slope ($h(i-1)+1=h(i)=h(i+1)-1$).
This point separates the walk $b$ into a left part $L$ and a right part $R$.
Let us reflect $R$ and paste it again to the right end of $L$.
This produces a walk $a'\in W_{2n-k}^{(k)}$.
As before, if $h(i+1)-1=h(i)=m+k$, we mark the point $i$, which will be
an end of strip (in the eventually reflected walk).
Otherwise, $h(i+1)=h(i)-1$, and we migrate the mark to the point
$i'=max\{ j<i | h(j+1)=h(j)-1=m+k\}$).
Let us now mark (by black dots) the {\it rightmost} 
intersections between $a'$
and the lines $h=k$, $h=k-1$, ..., $h=1$ at ascending slopes of $a'$,
and also the left end $(i=0,h=0)$ of $a'$.
We reconstruct the $k+1$ separations of floors using the following
rules (inverse of \rulamput)
\eqn\amputrul{\eqalign{
(1)\ \ \figbox{2.cm}{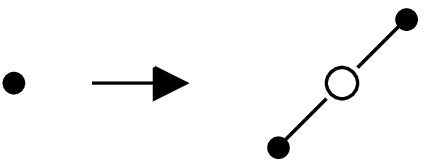} &\ (h\to h+2, o\to o+2) \cr
(2)\ \ \figbox{2.cm}{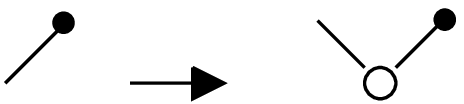} &\ (h\to h-1, o\to o+1) \cr
(3)\ \ \figbox{2.cm}{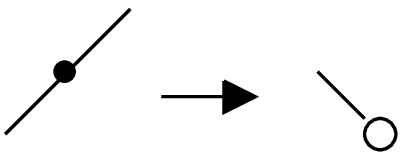} &\ (h\to h-3, o\to o+1) \cr} }
The corresponding separations have been represented
by empty circles. They all lie at height $h=1$ in the resulting
final walk $a''$. 
The three rules of \amputrul\ apply respectively 
(1) to the left end
of $a'$, (2) to
any of the $k-1$ intermediate points of intersection
with the lines $h=1$, ..., $h=k-1$, and (3)
to the rightmost intersection with the line $h=k$. 
The rules \amputrul\ therefore
result in a change of final height and order
$(h=k,o=2n-k)\to (h+2-(k-1)-3=0,o+2+(k-1)+1=2n+2)$, 
hence $a''\in W_{2n+2}$.
The last step consists simply in reflecting $a''$,
to produce $a={\bar{a''}}$, with a marked top of strip of 
height $m+k-k=m$ above the leftmost ($F_0$) floor,
and $a$ has a total of $k+1$ floors, all of height 0.
As before, the bijectivity of the map 
follows from the fact that we considered 
rightmost 
points of intersection, which makes the construction unique.
This bijection yields the number $c_{2n-k,2m+k}$ of walks 
in $W_{2n+2}$ with $k+1$ floors of height 0, and with a marked
top of strip of length $m$
above $F_0$. This is the first term of \numread.
\par
\noindent{\bf CASE (ii) :} 
We start from a walk
$a\in W_{2n+2}$, with $k+1$ floors $F_0$, $F_1$,...$F_k$, all of height
zero,
and with a marked top of strip of length $m$ above
its rightmost floor $F_k$.
By definition, we necessarily have $m\geq 2$, and in fact there
is one and only one strip of length 2 above the floor $F_k$ (the one
just above the right floor-end), and all other strips have 
length $\geq 3$.
We now construct a bijection between these walks and the $b\in W_{2n+2}$
with $k+1$ floors $F_0'$, $F_1'$, ..., $F_k'$,
all of height 0, and with a marked top of strip
of length $m-2$ above their leftmost floor $F_0'$.
\par
If $m=2$, the above remark shows that the number of walks $a$ 
with $k+1$ floors of height 0, and with a
marked top of strip of length $2$ above $F_k$ is equal to the
number of such walks, without marked top of strip (there is exactly
one such strip of length 2 per walk). 
Skipping the cutting-reflecting-pasting
procedure of the case (i) (we have no more marked top of strip),
we can still apply the amputation rules \rulamput\
on the walk $b'=\bar a\in W_{2n+2}$: this results in a walk 
$b\in W_{2n-k}^{(k)}$. Conversely, starting from any 
$b\in W_{2n-k}^{(k)}$, let us apply to it the inverse of 
the amputation rules \amputrul, after marking the rightmost
intersections at ascending slopes with the lines 
$h=k$, $h=k-1$, ..., $h=1$. This produces a walk $a'\in W_{2n+2}$,
and finally $a=\bar{a'}\in W_{2n+2}$ has $k+1$ floors of height 0.
This bijection yields the number $c_{2n-k,k}$ of walks
in $W_{2n+2}$ with $k+1$ floors of height 0, and a marked top
of strip of height $m=2$ above $F_k$. Together with the $k$-fold
cyclic degeneracy of the case (ii) this gives the second 
term of \numread, for $m=2$.
\par
\fig{The exchange map on walk diagrams of $W_{n}^{(h)}$, 
maps the
walks with a marked strip of length $m$ 
above their rightmost floor onto those with a marked strip
of length $m-2$ above the leftmost floor
(the corresponding strip of length $m=3$ is marked with a black dot
on the figure). We have indicated by a
thick broken line the portions exchanged. The double-layer of white boxes
on the rightmost floor is adapted to fit the exchange.}{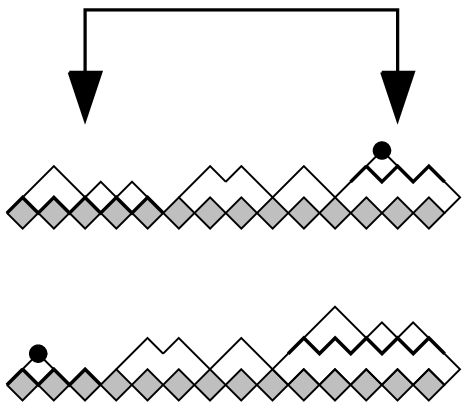}{6.cm}
\figlabel\exch
If $m\geq 3$, we simply {\it exchange} the floors $F_0$ and $F_k$ in the
following way. The floor $F_k$ is by definition topped by at least
two layers of white boxes (see Fig.\exch). 
Let $i_k,i_k+1,...,i_{k+1}$ denote the positions
occupied by $F_k$, the ends $i_k$ and $i_{k+1}$ being at
height 1. Let us cut out the portion $a_k$ of $a$ inbetween the
positions $i_k+2$ and $i_{k+1}-2$, both at height 3 (the level of the
second layer of white boxes).
Let us also cut the portion $a_0$ of $a$ above 
the leftmost floor $F_0$, inbetween the positions $i_0=0$ and $i_1-1$,
both at height 0. We form a walk $b\in W_{2n+2}$ by simply exchanging
the portions $a_0$ and $a_k$ in $a$, as depicted in Fig.\exch.
The marked top of strip on $a_k$ has been therefore transferred
above the leftmost floor of $b$, but as two layers of white 
boxes have been suppressed, all the lengths of strips have been 
decreased by 2. Hence the walk $b\in W_{2n+2}$ has $k+1$ floors
of height 0, and a marked top of strip of length $m-2$
above its leftmost floor $F_0'$. 
This construction is clearly invertible, by just exchanging again $a_k$
and $a_0$.
From the case (i) above, we learn
that the walk $b$ can be mapped onto  an  element
of $W_{2n-k}^{(2(m-2)+k)}=W_{2n-k}^{(2m+k-4)}$, in a bijective way.
This bijection yields the number $c_{2n-k,2m+k-4}$ of walks 
$a\in W_{2n+2}$ with $k+1$ floors of height 0, and with a marked top 
of strip of length $m$ above its rightmost floor $F_k$. 
With the overall $k$-fold cyclic degeneracy
mentioned above, this gives the second term 
in \numread\ for all $m\geq 3$.
\par
The mappings of the cases (i) and (ii) above complete the proof
of proposition 5, with the understanding that the case $m=1$
only gives rise to the case (i), hence the different answer.
\par
\subsec{The semi-meander determinant: the final formula}
Combining the results of propositions 4 and 5, namely eqs.\profour\
and \numread, we get the following formula for the numbers
$s_{n,m}^{(h)}$ of walk diagrams in $W_{n}^{(h)}$ with a marked
top of strip of length $m$ above its floor
\eqn\calsnmh{\eqalign{
s_{n,m}^{(h)}~&=~\sum_{k\geq 0} {h+k-1\choose k}
(c_{n-h-k,2m+k} +k c_{n-h-k,2m+k-4}) \qquad {\rm for} \ m\geq 2\cr
s_{n,1}^{(h)}~&=~\sum_{k \geq 0} {h+k-1\choose k}
c_{n-h-k,k+2} \qquad {\rm for}\ m=1\cr}  }
This is valid for $h\leq n-1$. If 
$h=n$, \fzer\  yields $s_{n,m}^{(n)}=0$ for all $m$.
By a direct calculation, we find
\par
\deb
\noindent{\bf PROPOSITION 6 :}
\par
The numbers of walks in $W_{2n+2}$ with $k+1$ floors of height 0 and
a marked end of strip of length $m$ read
\eqn\procalfi{\eqalign{
s_{n,m}^{(h)}~&=~ c_{n,h+2m} + h c_{n,h+2m-2} \qquad {\rm for} \ 
m\geq 2, \ h \leq n-1 \cr
s_{n,1}^{(h)}~&=~ c_{n,h+2} \qquad {\rm for} \ m=1, \ h \leq n-1\cr
s_{n,m}^{(n)}~&=~ 0 \qquad {\rm for} \ h=n \ {\rm and} \ {\rm all}\
m \geq 1\ cr}}
\fin
\par
The proof relies on the following classical
identity for binomial coefficients
\eqn\idbin{ \sum_{k=b-a}^{c-d} {k+a \choose b} {c-k \choose d} ~=~
{a+c+1 \choose b+d+1} }
for all integers $a$, $b$, $c$, $d$.
This is easily proved by use of generating functions. 
We now simply have to apply \idbin\ to the various sums
appearing on the r.h.s. of \calsnmh\
\eqn\varsums{\eqalign{
\sum_{k\geq 0} {k+h-1 \choose h-1} {n-h-k \choose {n-h\over 2}+m}
~&=~ {n \choose {n+h\over 2} +m} ~=~{n \choose {n-(h+2m)\over 2}}\cr
\sum_{k\geq 0} {k+h-1 \choose h-1} {n-h-k \choose {n-h\over 2}+m+1}
~&=~ {n \choose {n+h\over 2} +m+1} ~=~{n \choose {n-(h+2m)\over 2}-1}\cr}}
hence 
\eqn\firstpart{ \sum_{k\geq 0} {k+h-1 \choose k} c_{n-h-k,k+2m}~=~
c_{n,h+2m} }
and, noting that $k{k+h-1\choose k}=h{k+h-1\choose k-1}$, we also get
\eqn\secpart{ \sum_{k\geq 0} k {k+h-1\choose k} c_{n-h-k,k+2m-4}~=~
h c_{n,h+2m-2} }
The propositions 6 follows from \firstpart\ and \secpart.
\par
Substituting the result \procalfi\ above
into \recsdet, we finally get the semi-meander
determinant
\eqn\sdefinal{ \det {\cal G}_n^{(h)}(q) ~=~
\prod_{m=1}^{{n-h \over 2}+1} 
\big[\mu_m\big]^{-(c_{n,h+2m}+h c_{n,h+2m-2})} }
where we have absorbed the prefactor $\mu_1^{-hc_{n,h}}$
of \recsdet\ into the $m=1$ term of the product.
Finally, using the fact that $\mu_m=U_{m-1}(q)/U_{m}(q)$, for $m\geq 1$,
the theorem 2 follows. 
\par
\newsec{Conclusion}
\par
In this paper, we have
proved two determinant formulas for meanders and semi-meanders.
This has been done by the Gram-Schmidt orthogonalization 
of the corresponding bases 1 of the Temperley-Lieb algebra
or some of its subspaces.
The main philosophy of the construction leading to
the Gram-Schmidt orthogonalization of these bases 1
lies in the concept
of {\it box addition}, the building block of the definition of
the bases 2 elements. We believe that this type of construction should be
much more general and apply to many other cases of 
algebra-related Gram-Schmidt 
orthogonalization.
\par
An important remark about Theorems 1 and 2 above is that they
implicitely give the structure (including multiplicity) of the
zeros of the Gram determinants, considered as functions
of the variable $q$. Due to the definition of the Chebishev
polynomials, the zeros of the Gram determinant
always take the form
\eqn\zerconc{ q~=~ 2\ \cos ( {\pi} {m \over p+1} ) }
with $1 \leq m \leq p \leq {n-h \over 2}+1$ in the semi-meander case.
These zeros actually correspond to the cases when the corresponding
subspace of the Temperley-Lieb algebra is {\it reducible}
(there are linear combinations of the basis 1 elements which are
orthogonal
to all the basis 1 elements: i.e. there may be vanishing linear
combinations of the basis 1 elements, the basis 1 being therefore
no longer a basis at these values of $q$).
The multiplicity of these zeros is linked to the degree of reducibility
(namely to how many such independent linear combinations exist).
\par
Unfortunately, the information we obtain from these 
determinant formulas on the
meanders and the semi-meanders themselves is very difficult to exploit.
Indeed, quantities such as asymptotics 
(for large order) of the meander and semi-meander
numbers and distributions are only indirectly related to the 
Gram determinants, as they 
would rather involve the exact knowledge of the asymptotics of the
Gram matrices, or at least of their eigenvalue spectra.
However, the exact orthogonalization performed above is 
useful to derive new asymptotic formulas for the meander numbers, as
sums over walk diagrams (see \DGGB\ for the meander example).
We hope to return to this question in a later publication.
\par
The Theorems 1 and 2 above can probably be generalized in many
directions. A first possibility relies on the fact that
there exists a canonical Temperley-Lieb algebra attached to any 
non-oriented, connected graph (see
\DIF\ and references therein), which may still be interpreted
as the image of a walk diagram {\it on that graph}. 
The only constraint is that the number $q$ must be an eigenvalue
of the adjacency matrix of the graph (a matrix $G_{a,b}$
made of 1's and 0's
according to wheter the couple $(a,b)$ of vertices of the graph is
joined by a link or not).
In the examples
treated here, this graph is simply the set of heights, namely
the integer points on the (infinite) 
half-line, linked by segments between
consecutive pairs (hence $G_{i,j}=\delta_{j,i+1}+\delta_{j,i-1}$
for $i,j>0$ and $G_{0,j}=\delta_{j,1}$, with the eigenvalue $q$
for the (infinite) eigenvector $\vec{v}=(U_0(q),U_1(q),U_2(q),...)$).
But nothing prevents us from
considering more complicated graphs. 
We believe that there exists a
general determinant formula, associated to each such graph, expressing
the result in terms of features of the graph only (with 
$c_{2n,2m}$ replaced by a corresponding number of paths of given
length and given origin and end on the graph, and $U_m(q)$
replaced by the components of the
eigenvector of the adjacency matrix
for the eigenvalue $q$). 
\par
Another direction of generalization has to do with replacing the
Temperley-Lieb algebra by a more general quotient of the Hecke
algebra. Indeed, recall that the Temperley-Lieb algebra
$TL_n(q)$ is nothing but a simple quotient of the Hecke algebra,
defined as follows. The Hecke algebra $H_n(q)$ is defined by
generators $1, e_1, e_2,...,e_{n-1}$ satisfying the following
relations
\eqn\relaheck{ \eqalign{
(i)\ e_i^2~&=~ q \ e_i \cr
(ii)\ [e_i,e_j]~&=~0 \quad {\rm if} \ |i-j|>1 \cr
(iii) \ e_i e_{i+1} e_i -e_i~&=~e_{i+1} e_i e_{i+1} -e_{i+1} \cr}}
hence the Temperley-Lieb algebra is the quotient of the Hecke
algebra by the ideal generated by the elements
$e_i e_{i\pm 1} e_i -e_i$.   
This quotient was identified as the {\it commutant} of the quantum
enveloping algebra $U_{\hat q}(sl_2)$ 
acting on the fundamental representation
of $H_n(q)$, with $q={\hat q}+{\hat q}^{-1}$. 
More quotients are found by considering the
commutants of other quantum enveloping algebras 
(such as $U_{\hat q}(sl_k)$ for instance) \RESHE. 
These quotients await a good combinatorial interpretation, but
should lead to a natural generalization of meanders and semi-meanders.
We believe that many Gram determinants can still be computed exactly
in this framework.   
\par
\listrefs
\bye